 \ifx\compilefullpaper\undefined  
 \documentclass[pra,longbibliography,twocolumn,showpacs,nofootinbib,
                superscriptaddress,notitlepage]{revtex4-2}
 
 \pdfoutput=1 
 
 \usepackage{mathtools}
 \usepackage{amssymb}
 \usepackage{bm}
 \usepackage{amsthm}
 \usepackage{mathrsfs}
 \usepackage{stmaryrd}
 \usepackage{latexsym}
 
 \usepackage[table]{xcolor}   
 \renewcommand{\arraystretch}{1.2}
 \usepackage{booktabs}
 \usepackage{array,makecell}
 \usepackage{boldline}
 
 \usepackage[utf8]{inputenc}  
 \usepackage[protrusion=true,expansion=true]{microtype}
 \usepackage{ragged2e}
 \usepackage{setspace}
 \usepackage{lipsum}
 \usepackage{ulem}
 
 \usepackage{graphicx}
 \usepackage{float}
 \usepackage{verbatim}
 \usepackage[caption=false]{subfig}
 
 \usepackage{enumitem} 
 
 \usepackage{dsfont}
 \usepackage{url}      
 \usepackage{keyval}
 
 \usepackage[noend]{algpseudocode} 
 \usepackage{algorithm}

 \definecolor{cardinal}{rgb}{0.827,0,0}

 \usepackage[colorlinks=true,hyperindex,breaklinks,
             linkcolor=blue,urlcolor=blue,citecolor=blue]{hyperref}
 \usepackage[capitalise]{cleveref}
 
\begin{document}
 \title{Fast and accurate AI-based pre-decoders for color codes}

\author{Jan Olle}
\affiliation{NVIDIA Corporation, USA}

\author{Christopher Chamberland}
\thanks{Corresponding author}
\email{cchamberland@nvidia.com}
\affiliation{NVIDIA Corporation, USA}

\author{Muyuan Li}
\affiliation{NVIDIA Corporation, USA} 

\author{Igor Baratta}
\affiliation{NVIDIA Corporation, USA} 
 
\begin{abstract}
Color codes are promising alternatives to surface codes for universal fault-tolerant quantum computing due to their simpler lattice-surgery protocols and the transversal implementation of logical Clifford gates. However, their practical deployment has been limited by slower decoding algorithms and worse logical failure rates and thresholds compared to surface codes. Although AI-based logical-flip decoders have recently been proposed to address these challenges, no clear framework currently exists for implementing such decoders within the parallel block-wise decoding schemes in both space and time required for large-scale fault-tolerant computation. AI-based pre-decoders offer a scalable alternative due to their local nature. By performing spacelike corrections on physical qubits and timelike corrections on stabilizer measurements, pre-decoders are naturally compatible with parallel block-wise decoding schemes and lattice-surgery protocols. In this work, we introduce AI-based pre-decoders for triangular color codes. We present a novel neural-network architecture for their implementation and develop methods to simplify the complex training data generated by color-code syndrome-extraction circuits containing feedforward operations. Remarkably, we find that both logical failure rates (LERs) and runtimes improve relative to raw Chromobius decoding as the code distance increases. For example, at code distance d=31 and physical error rate $p=0.3\%$, our pre-decoder + Chromobius pipeline improves the logical failure rate by a factor of 347x while reducing runtime by 7.33x compared to raw Chromobius decoding alone. These results demonstrate that AI-based pre-decoding can substantially narrow the performance gap between color codes and surface codes, bringing color codes closer to practical large-scale fault-tolerant quantum computation.
\par\medskip
\noindent\centering\small
\textbf{Code:} \href{https://github.com/nvidia/ising-decoding}{GitHub}
\qquad
\textbf{Models:} \href{https://huggingface.co/collections/nvidia/nvidia-ising}{Hugging Face}
\par
\end{abstract}
 
 \maketitle
 
\section{Introduction}
\label{sec:Intro}

Fault-tolerant quantum error correction will be essential for the implementation of large-scale quantum algorithms. Numerous quantum error-correcting codes have been proposed as candidates for universal fault-tolerant quantum computation on realistic hardware platforms. Among the most prominent are planar surface codes \cite{DennisSurface,fowler2012surface,TomitaRotatedSurface2014} and color codes \cite{BombinTopoDistill,KubicaColor2015,ChamberlandColor2020,GidneyColor2023}. Both are two-dimensional topological codes whose stabilizers can be measured using only nearest-neighbor interactions. Moreover, universal fault-tolerant logic can be implemented using nearest-neighbor operations when lattice surgery \cite{fowler2018low,Litinski2018latticesurgery,Litinski19magic,Litinski19,Chamberland22,Chamberland22b} is combined with magic-state distillation protocols \cite{BravyiKitaevMagic,Campbell2018magicstateparity,Litinski19,Gidney19,Chamberland2019Magic1,chamberland2020very,Gidney_cultivation}.

Color codes possess several advantages over surface codes for fault-tolerant logic. Most notably, all logical Clifford gates admit transversal implementations, substantially simplifying logical gate constructions. In addition, logical $Y$-measurements can be implemented without the use of twist defects and the asymmetric routing-space corridor configurations required in surface-code architectures.

Despite these advantages, surface codes remain the dominant architecture currently considered for large-scale fault-tolerant quantum computation. Two primary factors contribute to this. First, color codes are significantly more difficult to decode. State-of-the-art color-code decoders such as Chromobius \cite{GidneyColor2023} are generally slower than Minimum-Weight Perfect Matching (MWPM) \cite{Edmonds_1965,HiggottPyMatch} and Union-Find (UF) \cite{DelfosseUnionFind} decoders commonly used for surface codes. This makes real-time decoding more challenging, particularly for superconducting hardware architectures where decoding latencies approaching $1 \mu \text{s}$ per syndrome round may be required to avoid the exponential backlog problem \cite{TerhalBacklog,ChambsLocalNN22,CampbellParallelV1,AlibabaParallel}. Second, color codes typically require more physical qubits than surface codes to achieve the same target logical error rate (LER), resulting in larger resource overheads.

Recently, several new decoding approaches have been proposed to address these limitations. Ref.~\cite{VibeLSD} introduced the VibeLSD decoder, which achieves target LERs using a number of physical qubits comparable to that of surface codes. However, because the algorithm relies on serialized belief propagation, further work is needed to demonstrate decoding runtimes compatible with real-time operation at large code distances. AI-based decoders for color codes such as AlphaQubit 2 \cite{AlphaQubit2} achieve near-optimal LER performance for both surface and color codes when provided with sufficient training data and optimization time. However, it remains unclear how such logical-classifier decoders can be integrated into the parallel block-wise decoding schemes in both space and time required for large-scale lattice-surgery computations \cite{CampbellParallelV1,AlibabaParallel}.

In lattice-surgery protocols, logical operations often produce extremely large merged code patches whose associated space-time decoding volumes are too large to decode monolithically in real time. Parallel block-wise decoding methods address this challenge by partitioning the global decoding volume into commit regions, associated buffer regions, and cleanup regions. To resolve residual syndromes within cleanup regions, physical spacelike and timelike corrections must be applied within commit regions. Since logical-classifier decoders infer only the values of logical observables rather than physical corrections throughout the decoding volume, it is unclear how they can be directly adapted to resolve cleanup-region syndromes within such frameworks.

In this work, we propose an AI-based pre-decoder for triangular color codes. Pre-decoders \cite{Gicev2023scalablefast,ChambsLocalNN22,Australia3DConvPred,COLTI_Pre_surface} are trained to apply local corrections using only local spatio-temporal information. They output both spacelike and timelike physical corrections across all syndrome-measurement rounds and can be trained on a fixed volume $d_x \times d_z \times d_m$ while generalizing to decoding volumes of arbitrary size $d'_x \times d'_z \times d'_m$. Because pre-decoders are local and probabilistic, residual errors generally remain after correction and must therefore be resolved using a global decoder. However, since the runtime of most global decoders depends strongly on syndrome density, pre-decoders can substantially accelerate global decoding by sparsifying the residual syndrome. If the pre-decoder latency is sufficiently small, the combined pre-decoder + global decoder pipeline can achieve lower total runtime than the standalone global decoder. In many cases, such pipelines also improve logical error rates relative to the global decoder alone. Most importantly, because pre-decoders generate physical spacelike and timelike corrections throughout the decoding volume, they are naturally compatible with parallel block-wise decoding schemes in both space and time.

In this work, we use Chromobius \cite{GidneyColor2023} as the global decoder due to its strong combination of decoding performance and runtime efficiency. Although VibeLSD achieves lower LERs than Chromobius, competitive real-time decoding runtimes have not yet been demonstrated. We show that our pre-decoder + Chromobius pipeline achieves both substantial LER reductions and runtime improvements relative to standalone Chromobius decoding. Moreover, these improvements become increasingly pronounced as the code distance grows and persist both above and below the physically relevant regime of $p = 0.1\%$. For example, at $d=31$ and $p=0.3\%$, our pipeline achieves a 347x reduction in LER together with a 7.33x runtime improvement relative to Chromobius alone.

This paper is organized as follows. In \cref{sec:ColorCodeReview}, we briefly review the fundamentals of color codes. In \cref{sec:NNArchHyperParam}, we describe the neural-network architecture used for our AI-based pre-decoders and introduce data-processing techniques that substantially improve training performance. In \cref{sec:Numerics}, we present detailed numerical results for both LER and runtime improvements obtained using the pre-decoder + Chromobius pipeline. Finally, we conclude and discuss future directions in \cref{sec:Conclusion}.

\section{Brief review of the color code}
\label{sec:ColorCodeReview}

\begin{figure*}
     \centering
 \subfloat[\label{fig:ColorCodeLattice} ]{\includegraphics[width=.5\textwidth]{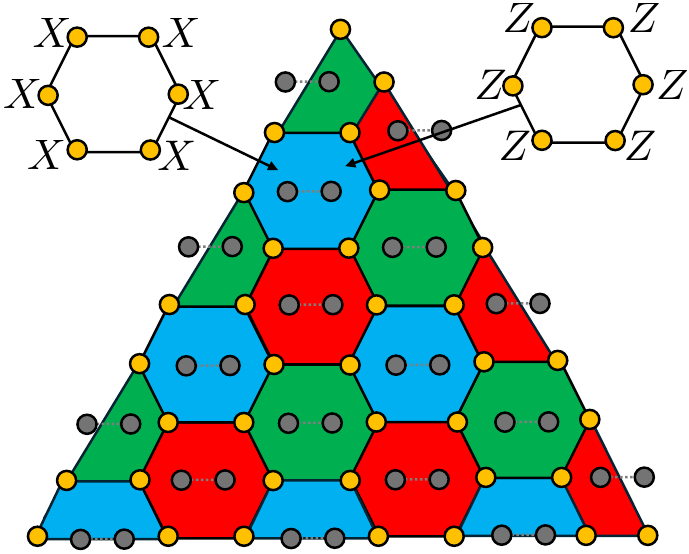}}
 \vfill
 \subfloat[\label{fig:Circuits_W6} ]{\includegraphics[width=.5\textwidth]{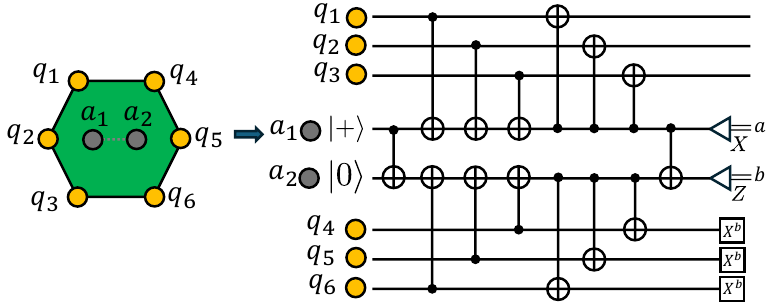}}
 \subfloat[\label{fig:Circuits_W4} ]{\includegraphics[width=.5\textwidth]{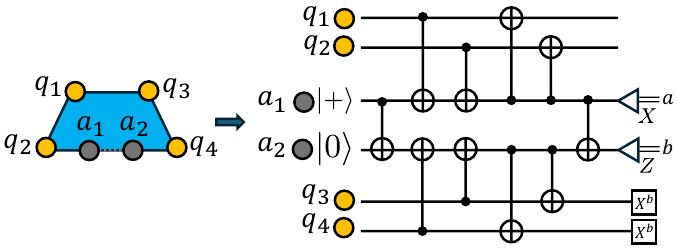}}
 \vfill
\subfloat[\label{fig:Boundary_W4} ]{\includegraphics[width=.5\textwidth]{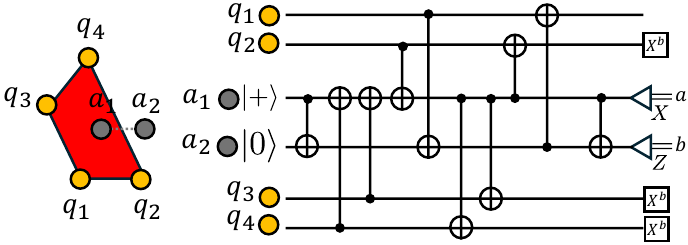}}
\subfloat[\label{fig:Boundary_W4_V2} ]{\includegraphics[width=.5\textwidth]{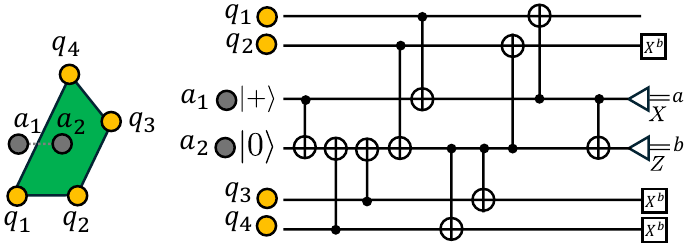}}
     \caption{ (a) A $d=7$ lattice for the triangular color code. Each weight-4 and weight-6 plaquette corresponds to both an $X$ and $Z$-type stabilizer. Data qubits shown by yellow vertices, with gray vertices corresponding to the ancillas used to measure the stabilizers. Logical representatives for $X_L$, $Z_L$ and $Y_L$ operators correspond to tensor product of $X$, $Z$ and $Y$ Paulis on data qubits along any of the boundaries of the triangle. (b) Circuit to measure both weight-6 $X$ and $Z$-type stabilizers. (c) Same as (b) but for weight-4 stabilizers. In (d) and (e), we show the circuits used for red and green weight-four stabilizers which maintains the regular grid tilling of all qubits.  }
     \label{fig:Color_Code_Review}
\end{figure*}

In this section, we review the key properties of triangular color codes relevant to the pre-decoder architecture described in \cref{sec:NNArchHyperParam}.

Triangular color codes are obtained by cutting a tiling of hexagons with an equilateral triangle, resulting in the lattice shown in \cref{fig:ColorCodeLattice}. The resulting lattice satisfies the 3-valence condition: all vertices, except for the three corner vertices, are incident to exactly three edges. The lattice is also 3-colorable, meaning that each plaquette can be assigned one of three colors (e.g., red, green, and blue) such that neighboring plaquettes sharing an edge have different colors \cite{BombínColor2015,KubicaColor2015,ChamberlandColor2020}.

Each plaquette supports both an X-type and a Z-type stabilizer acting on the data qubits located at its vertices (shown as yellow vertices in \cref{fig:ColorCodeLattice}). For each Pauli type $P \in \{X,Z\}$, the logical operator $P_L$ can be represented by a string of identical Pauli operators supported along any one of the three boundaries of the triangular lattice.

In this work, we use the ``superdense’’ syndrome-extraction circuits introduced in Ref.~\cite{GidneyColor2023} and shown in \cref{fig:Circuits_W6,fig:Circuits_W4} to measure the weight-6 and weight-4 stabilizers of the color code. To maintain a regular lattice geometry for both data qubits and ancillas, we use the modified boundary circuits shown in \cref{fig:Boundary_W4,fig:Boundary_W4_V2} for the red and green weight-4 stabilizers.

A key advantage of the circuits in \cref{fig:Color_Code_Review} is that the $X$- and $Z$-type stabilizers can be measured simultaneously. Specifically, the ancilla prepared in $| + \rangle$ ($| 0 \rangle$) encodes the $X$-type ($Z$-type) stabilizer measurement outcome. Similar circuits were previously considered in Ref.~\cite{BaireutherColor2019}, where the second ancilla was instead used as a flag qubit \cite{ChaoFlag1,Chamberland2018flagfaulttolerant,ChaoFlag3,Chamberland2019Magic1,chamberland2020very}. However, that approach requires repeating the circuit twice in order to measure both stabilizers, resulting in a larger number of fault locations.

Recently, Ref.~\cite{GidneyColor2023} introduced Chromobius, an open-source implementation of the m\"{o}bius color-code decoder. Chromobius relies on Minimum-Weight Perfect Matching (MWPM) \cite{Edmonds_1965,HiggottPyMatch} applied to subgraphs of the decoding problem, and consequently its runtime depends strongly on the syndrome density (i.e., the number of detection events). As such, Chromobius is particularly well suited for use within an AI-based pre-decoding pipeline \cite{Gicev2023scalablefast,ChambsLocalNN22,Australia3DConvPred,COLTI_Pre_surface}, where the pre-decoder reduces the syndrome density prior to global decoding and thereby simplifies the downstream matching problem. Nevertheless, we emphasize that any global decoder can be used in conjunction with our AI-based pre-decoder.
 
\section{Neural network architecture and hyperparameters}
\label{sec:NNArchHyperParam}

\begin{figure}
     \centering
 \subfloat[\label{fig:PreDecOverviewV1} ]{\includegraphics[width=.4\textwidth]{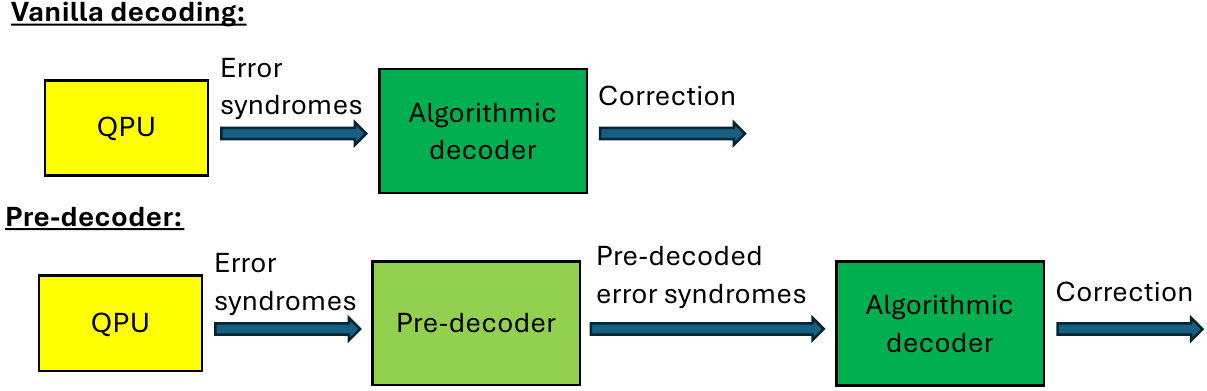}}
 \caption{Comparison between a vanilla decoding algorithm and a decoding algorithm that uses a pre-decoder to pre-process the error syndromes. When using a pre-decoder, the pre-decoder receives the error syndrome from the QPU and applies spacelike and timelike corrections across all syndrome measurement rounds that were used as inputs. The new error syndrome obtained from the corrections are then passed to an algorithmic decoder to apply the final set of corrections. Figure taken from Ref.~\cite{COLTI_Pre_surface}. }
 \label{fig:PreDecOverview}
\end{figure}

\begin{figure*}
     \centering
 \subfloat[\label{fig:ConvArchEX} ]{\includegraphics[width=.6\textwidth]{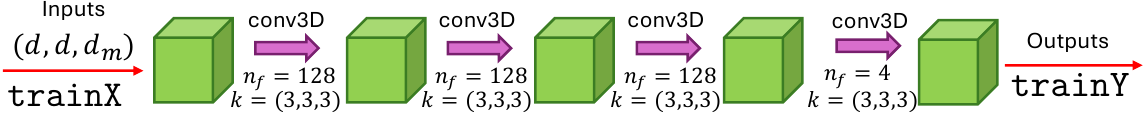}}
 \vfill
  \subfloat[\label{fig:Arch_Cascade} ]{\includegraphics[width=.8\textwidth]{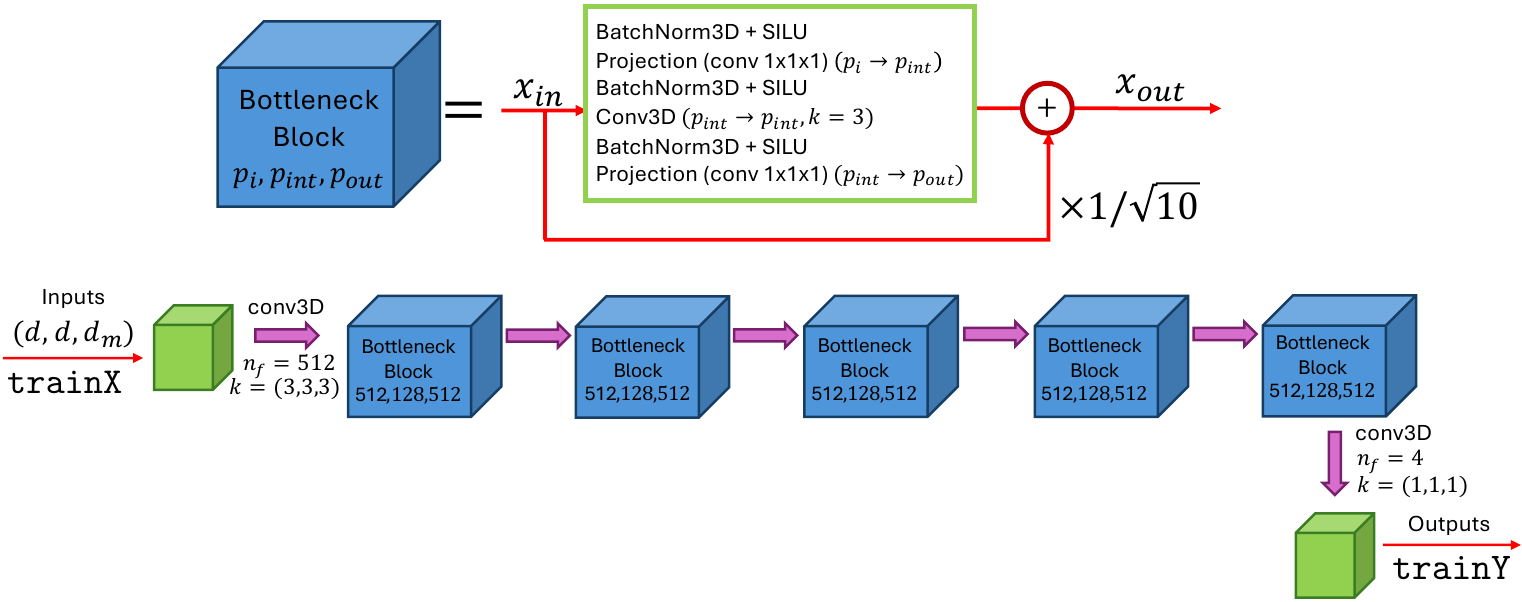}}
 \caption{ (a) Illustration of the fully convolutional neural network architecture used to train our pre-decoders. The input data is labeled \texttt{trainX} and the output data labeled \texttt{trainY}. We use three-dimensional convolutional layers which can have varying kernel sizes and number of filters. The output of the last convolution layer goes through a sigmoid activation function in order to generate the spacelike and timelike predictions. Figure taken from Ref.~\cite{COLTI_Pre_surface}. (b) Alternative architecture which uses Bottleneck blocks taken from Refs.~\cite{7780459,Cascade_decoder}. We find that such architectures achieves lower LERs while using less parameters compared to the fully three-dimensional convolutional architecture with the same number of layers, conv3D kernel sizes and filters. }
 \label{fig:ConvArch}
\end{figure*}

In this section, we present the AI-based pre-decoder architecture used for triangular color codes. Many of the techniques introduced here are adapted from Ref.~\cite{COLTI_Pre_surface} and generalized to the color-code setting. An overview of the full decoding pipeline incorporating a pre-decoder is shown in \cref{fig:PreDecOverview}.

At a high level, the pre-decoder receives syndrome data from the quantum processing unit (QPU) and applies local spacelike and timelike corrections. Because AI-based pre-decoders are probabilistic and local in nature, residual errors may remain after correction. Processed syndromes are therefore computed from the residual detection events together with the current values of the relevant logical observables. These processed syndromes are subsequently passed to a global decoder, which computes the final logical-observable corrections. In this work, Chromobius is used as the global decoder.

In \cref{subsecd:InputtrainGridMap}, we review the fully convolutional neural-network architecture used for pre-decoding and introduce a mapping from the triangular color-code lattice onto a rectangular grid. Using this mapping, we describe how the input data used for both training and inference is generated. In \cref{subsecd:OutputTrain}, we explain how spacelike and timelike failures are labeled within the grid representation. We additionally review data-processing techniques, first introduced in Ref.~\cite{COLTI_Pre_surface}, which substantially reduce the number of residual timelike failures following pre-decoding.

In \cref{subsecd:RemoveFakeErrorDiff}, we describe an important simulation procedure used during training-data generation to eliminate artificial $X$-type error asymmetries arising from the feedforward operations present in the syndrome-extraction circuits of \cref{fig:Color_Code_Review}. Finally, in \cref{subsecd:ColorCodeHomological}, we introduce homological-equivalence transformations tailored to triangular color codes that add additional structure to the labeled training data. We find that the data processing methods of \cref{subsecd:RemoveFakeErrorDiff,subsecd:ColorCodeHomological} significantly improve pre-decoder performance. 

\subsection{Input training data and grid mapping}
\label{subsecd:InputtrainGridMap}

\begin{figure*}
     \centering
 \subfloat[\label{fig:GridMappingEX} ]{\includegraphics[width=.5\textwidth]{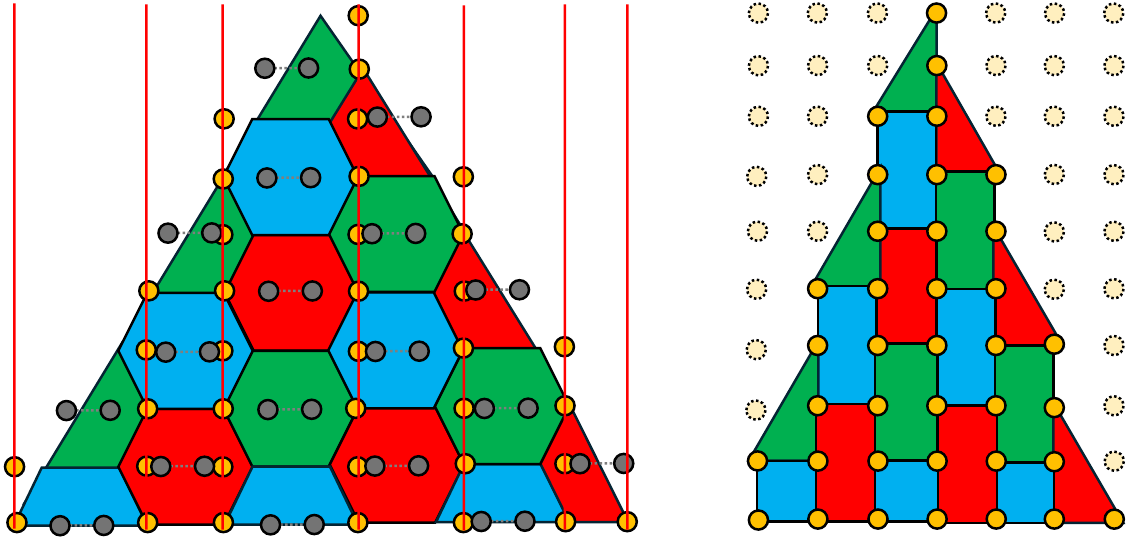}}
 \vfill
 \subfloat[\label{fig:MappingStabsToData} ]{\includegraphics[width=.5\textwidth]{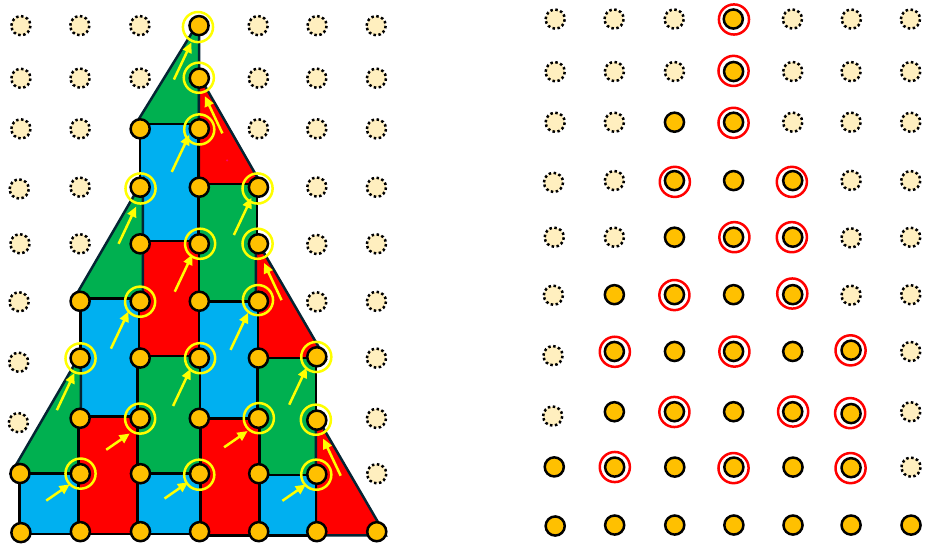}}
     \caption{ (a) Illustration of the mapping of the triangular color code to a two-dimensional grid. Vertical red lines are added to each data qubit at the bottom of the triangle. For each weight-4 and weight-6 stabilizer in between the red lines, data qubits closest to the right-most vertical line are mapped to that line. Similarly, data qubits closest to the left-most vertical line are mapped to that line. Such a mapping results in the figure on the right, resulting in rectangular weight-6 stabilizers and square or triangular weight-4 stabilizers. (b) Illustration of the mapping of stabilizer measurements to data qubit positions. A data qubit circled in red indicates that a stabilizer measurement is mapped to that data qubit. }
     \label{fig:GridMapping}
\end{figure*}

In this work we consider two architectures for our AI-based pre-decoders. The first uses a fully convolutional neural-network architecture with three-dimensional convolutions which allows our networks to be applicable to input volumes which are of different size than the one it was trained on. An illustration of the architecture is shown in \cref{fig:ConvArchEX}. We also consider a similar architecture, but where the three-dimensional convolutions are replaced with Bottleneck blocks \cite{7780459,Cascade_decoder} shown in \cref{fig:Arch_Cascade}. We found that architectures using Bottleneck blocks result in lower LERs while using fewer parameters compared to the fully convolutional architectures shown in \cref{fig:GridMappingEX}. However despite the smaller parameter count, they come at a slower runtime cost compared to similar fully convolutional networks. 

The relevant hyperparameters of the architecture in \cref{fig:ConvArchEX} are the number of convolutional layers, the kernel sizes, and the number of filters in each layer. For the architecture in \cref{fig:Arch_Cascade}, we also have the $p_{int}$ parameter used in the projection layers as well as $p_{out}$ (note that we fix the scaling factor to $1 / \sqrt{2L}$ with $L=5$ used in the skip connections). These hyperparameters determine tradeoffs between pre-decoder latency and decoding accuracy, which are explored in \cref{sec:Numerics}. We denote the input data to the neural network by \texttt{trainX} and the labeled output data used during training by \texttt{trainY}. In this subsection, we describe the construction of \texttt{trainX}.

We first map stabilizer-measurement outcomes to a rectangular grid so that the syndrome data can be processed by convolutional layers. The mapping is performed in two steps. First, weight-6 and weight-4 stabilizers are embedded into a rectangular-grid representation, as illustrated in \cref{fig:GridMappingEX}. Second, each stabilizer measurement outcome is assigned to a data-qubit position on this grid, as shown in \cref{fig:MappingStabsToData}. For all stabilizers except the red weight-4 stabilizers, the measurement outcome is mapped to the data qubit in the top-right position of its support, for both $X$- and $Z$-type stabilizers. For red weight-4 stabilizers, the measurement outcome is instead mapped to the data qubit in the top-left position of its support. The data qubits circled in red in \cref{fig:MappingStabsToData} indicate the grid locations that receive stabilizer-measurement outcomes. If the corresponding outcome is -1, a value of 1 is inserted at that grid location; otherwise, the inserted value is 0. All other data-qubit locations are assigned the default value 0.

Let $\texttt{x\_type}(k,j)$ and $\texttt{z\_type}(k,j)$ denote the $X$- and $Z$-type detection-event grids, respectively, for the $k$-th syndrome-measurement round and the $j$-th shot. Detection events in round $k$ are obtained by summing the mapped stabilizer values from rounds $k-1$ and $k$ modulo 2. For a $D_x \times D_y$ rectangular grid, the first two channels of \texttt{trainX} are defined as
\begin{align}
     \texttt{trainX}(j,1{:}D_x,1{:}D_y,k,1) &= \texttt{x\_type}(k,j), \label{eq:encTrainX1} \\
     \texttt{trainX}(j,1{:}D_x,1{:}D_y,k,2) &= \texttt{z\_type}(k,j).
     \label{eq:encTrainX2}
\end{align}

Similarly to Refs.~\cite{ChambsLocalNN22,COLTI_Pre_surface}, we also encode geometric information about the lattice in \texttt{trainX}. These additional channels allow the network to distinguish between bulk and boundary syndrome statistics. We denote these channels by $\texttt{x\_present}$ and $\texttt{z\_present}$ for the $X$- and $Z$-type stabilizers, respectively. These two channels are identical except in the first and last syndrome-measurement rounds, where they depend on the basis in which the data qubits are prepared and measured, as explained below.

The construction of $\texttt{x\_present}$ and $\texttt{z\_present}$ uses the same grid mapping as in \cref{fig:MappingStabsToData}. However, instead of mapping a stabilizer-measurement outcome to the corresponding grid location, we map the normalized stabilizer weight. The normalized weight is obtained by dividing the stabilizer weight by the maximum stabilizer weight, which is six in this case, so that all values lie in the interval $[0,1]$. Grid locations that do not receive a stabilizer mapping, i.e., data qubits not circled in red in \cref{fig:MappingStabsToData}, are assigned value 0. For the example shown in \cref{fig:MappingStabsToData}, we have
\begin{align}
     \texttt{x\_present}(k) &= 
 \begin{bmatrix}
0 & 0 & 0 & 2/3 & 0 & 0 & 0 \\
0 & 0 & 0 & 2/3 & 0 & 0 & 0 \\
0 & 0 & 0 & 1 & 0 & 0 & 0 \\
0 & 0 & 2/3 & 0 & 1 & 0 & 0 \\
0 & 0 & 0 & 1 & 2/3 & 0 & 0 \\
0 & 0 & 1 & 0 & 1 & 0 & 0 \\
0 & 2/3 & 0 & 1 & 0 & 1 & 0 \\
0 & 0 & 1 & 0 & 1 & 2/3 & 0 \\
0 & 2/3 & 0 & 2/3 & 0 & 2/3 & 0 \\
0 & 0 & 0 & 0 & 0 & 0 & 0
 \end{bmatrix}
 \label{eq:xpresent}
\end{align}
for $1<k<d_m$, where $d_m$ is the total number of syndrome-measurement rounds. In this range, $\texttt{z\_present}(k)$ is identical to $\texttt{x\_present}(k)$.

To encode temporal boundary information associated with state preparation and measurement, we modify the geometric channels in the first and last rounds. If the data qubits are prepared in $|0\rangle$ and measured in the $Z$-basis in round $d_m$, we set $\texttt{x\_present}(1)$ and $\texttt{x\_present}(d_m)$ to all-zero matrices. Similarly, if the data qubits are prepared in $|+\rangle$ and measured in the $X$-basis in round $d_m$, we set $\texttt{z\_present}(1)$ and $\texttt{z\_present}(d_m)$ to all-zero matrices. The geometric channels are then encoded as
\begin{align}
 \texttt{trainX}(j,1{:}D_x,1{:}D_y,k,3) &= \texttt{x\_present}(k), \label{eq:encTrainX3} \\
 \texttt{trainX}(j,1{:}D_x,1{:}D_y,k,4) &= \texttt{z\_present}(k).
 \label{eq:encTrainX4}
\end{align}

\begin{figure*}
     \centering
 \subfloat[\label{fig:S1S2FigEX} ]{\includegraphics[width=.5\textwidth]{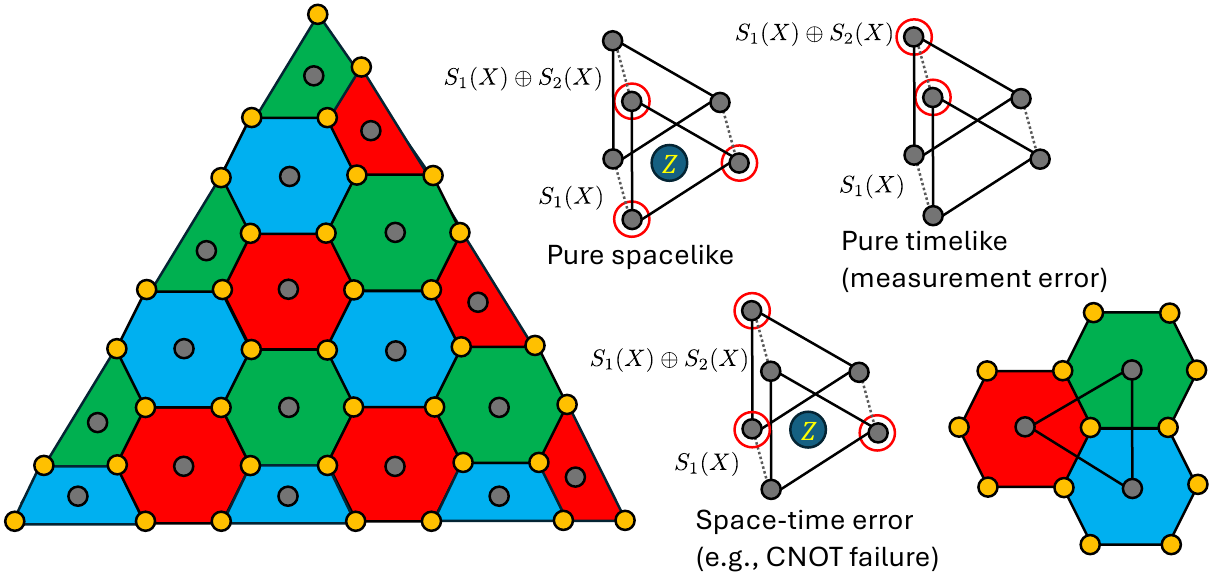}}
     \caption{ Example illustration of $S_1(X) \oplus S_2(X)$ computations used in \cref{Algo:TimelikeOutputGen} for various fault mechanisms. Only pure timelike failures (which arise from measurement errors) and space-time failures (for instance CNOT gate failures) result in a non-zero value of $S_1(X) \oplus S_2(X)$. The examples provided are for failures arising in the bulk of the color code lattice. The illustration uses a single ancilla for simplicity to focus on just $X$-type stabilizers, but the methods apply in the same way to the circuit in \cref{fig:Color_Code_Review}. }
     \label{fig:S1S2Fig}
\end{figure*}

\subsection{Output training data and data processing}
\label{subsecd:OutputTrain}

\begin{algorithm}[H]
 \caption{Timelike output channel generation}
 \begin{algorithmic}
 \For{$k = 1$ to $d_m - 1$}
     \State Let $E_k$ be the errors generated by the noise model at each fault location in syndrome measurement round $k$.
         \State Propagate $E_k$ and compute:
         \State \quad $X$ and $Z$ stabilizer syndromes $s_1(X)$, $s_1(Z)$
         \State Let $E^{(k)}_{\text{out}}$ be the output errors from propagating $E_k$.
         \State Propagate $E^{(k)}_{\text{out}}$ and compute:
         \State \quad $X$ and $Z$ stabilizer syndromes $s_2(X)$, $s_2(Z)$
         \State $\texttt{trainY}(j,1{:}D,1{:}D,k,3) \gets s_1(X) \oplus s_2(X)$
         \State $\texttt{trainY}(j,1{:}D,1{:}D,k,4) \gets s_1(Z) \oplus s_2(Z)$
 \EndFor
\end{algorithmic}
\label{Algo:TimelikeOutputGen}
\end{algorithm}

We follow Ref.~\cite{COLTI_Pre_surface} for generating the labeled output data \texttt{trainY}. As a first step, let $E^{(j)}(X)_{(i,k)} \in \{I,X\}$ for the $i$'th \textit{data} qubit in round $k$ and the $j$'th shot. We define the error difference in round $k$ as $\tilde{X}^{(j)}_{i,k} = E^{(j)}(X)_{i,k} \oplus E^{(j)}(X)_{i,k-1}$. We then write
 \begin{align}
     \tilde{X}^{(j)}_{k} \equiv (\tilde{X}^{(j)}_{(1,k)}, \cdots \tilde{X}^{(j)}_{(D_xD_y,k)}).
     \label{eq:XerrorDiff}
 \end{align}
 A similar definition for $\tilde{Z}^{(j)}_{k}$ is obtained for $Z$ errors, i.e.
 \begin{align}
     \tilde{Z}^{(j)}_{k} \equiv (\tilde{Z}^{(j)}_{(1,k)}, \cdots \tilde{Z}^{(j)}_{(D_xD_y,k)}).
     \label{eq:ZerrorDiff}
 \end{align}
 The first two channels of \texttt{trainY} are then given by
 \begin{align}
     \texttt{trainY}(j,1{:}D_x,1{:}D_y,k,1) &= \tilde{Z}^{(j)}_{k}, \label{eq:TrainY1} \\
     \texttt{trainY}(j,1{:}D_x,1{:}D_y,k,2) &= \tilde{X}^{(j)}_{k},
     \label{eq:TrainY2}
\end{align}
for the $j$-th shot and $k$-th training example. That is, the first two channels track changes in $Z$ and $X$-type Pauli errors obtained by generating faults at each location of the syndrome extraction circuits for the color code and propagating the errors to obtain final data qubit errors of the $D_x \times D_y$ grid shown in \cref{fig:GridMapping}. We note that special care is needed to track changes $\tilde{X}^{(j)}_{k}$ given the feedforward operations in the syndrome extraction circuits of \cref{fig:Color_Code_Review}. More details are given in \cref{subsecd:RemoveFakeErrorDiff}.

Next we consider the labels for timelike failures which are encoded in channels three and four of \texttt{trainY}. To extract timelike failures from arbitrary fault mechanisms, we use \cref{Algo:TimelikeOutputGen}.

An illustration of the computation of $s_1(X) \oplus s_2(X)$ adapted to the color code lattice is shown in \cref{fig:S1S2Fig}.

As was shown in Ref.~\cite{COLTI_Pre_surface}, the time steps at which spacelike and space-time errors occur is of particular importance when constructing the labels for \texttt{trainY}. For instance, suppose a $X$ error occurs in syndrome measurement round $1 < k < d_m$ during the implementation of the circuit in \cref{fig:Circuits_W6} in the bulk of the color code lattice. Such an error anti-commutes with three $Z$-type stabilizers. However if the error occurs at a time step where it only gets detected in round $k+1$, \texttt{trainY} would contain the $X$ error label in round $k$ with the associated syndrome given in \texttt{trainX} in round $k+1$. Such scenarios can result in our networks learning patterns that result in residual timelike failures when applying corrections during inference time. Such a problem can be avoided by treating the $X$ error as an input error in round $k+1$. More generally, we can use \cref{Algo:DataGenOptimize} as a data processing protocol when generating \texttt{trainX} and \texttt{trainY}.
\begin{algorithm}[H]
\caption{Data generation protocol}
\begin{algorithmic}
\State Let $e^{(0)} = \boldsymbol{0}$ be the empty error vector.
\For{$k = 1$ to $d_m - 1$}
     \State Initialize $e^{(k)} = \boldsymbol{0}$.
     \State Let $E_k$ be the full set of faults generated by the noise model at each fault location in syndrome measurement round $k$.
     \State Append $e^{(k-1)}$ to $E_k$.
     \State Let $N_{E_k}$ be the number of faults in $E_k$, and $e^{(k)}_j$ be the $j$-th fault in $E_k$ (with $1 \le j \le N_{E_k}$).
     \For{$j = 1$ to $N_{E_k}$}
         \State Propagate $e^{(k)}_j$ through the stabilizer measurement circuits of the color code.
         \State Let $s_{e^{(k)}_j}$ be the measured syndrome arising from the fault $e^{(k)}_j$.
         \State Let $|s_{e^{(k)}_j}|$ be the Hamming weight of $s_{e^{(k)}_j}$.
         \If{ $|s_{e^{(k)}_j}| > 0$}
             \State Update \texttt{trainX} and \texttt{trainY} as described in \cref{subsecd:InputtrainGridMap,subsecd:OutputTrain}.
         \Else
             \If{$e^{(k)}_j$ results in a non-trivial data qubit error $e^{(k)}_{d_j}$}
                 \State Append $e^{(k)}_{d_j}$ in time step 1 (so as to be treated as an input error) to $e^{(k)}$ and ignore updates to \texttt{trainY}.
             \EndIf
         \EndIf
     \EndFor
 \EndFor
\end{algorithmic}
\label{Algo:DataGenOptimize}
\end{algorithm}

\begin{table*}[htbp]
 \centering
 \begin{tabular}{|c|c|}
 \hline
 Data qubit on control of CNOT, ancilla on target & Data qubit on target of CNOT, ancilla on control \\
 \hline
 $YZ \to ZZ \oplus XI$ & $ZY \to ZZ \oplus IX$   \\
 $YX$ no decomposition necessary & $XY \to XX \oplus IZ$   \\
 $YY \to ZZ \oplus XX$ & $YY$ no decomposition necessary   \\
 \hline
 \end{tabular}
 \caption{Complete $Y$ error decomposition rules on two qubits following a CNOT gate. We use the format $P_1P_2$ where the Pauli $P_1$ is on the control of a CNOT and $P_2$ is on the target. The symbol $\oplus$ denotes a bookkeeping decomposition into separately propagated Pauli components whose contributions to the binary output labels are combined modulo two; it does not denote a physical tensor product or a probabilistic mixture.}
 \label{tab:Ydecomp}
\end{table*}
Additional care is required when analyzing faults containing $Y$ errors. For instance, the $X$ component of a single $Y$ error on a data qubit could be detected in round $k$, while the $Z$ component is detected in round $k+1$. For this reason, prior to applying \cref{Algo:DataGenOptimize}, we decompose the relevant $Y$-containing faults into separately propagated Pauli components. A single-qubit $Y$ error is represented as $X \oplus Z$, with the two components propagated independently. For two-qubit faults containing at least one $Y$ error, the decomposition must respect how the $X$ and $Z$ components propagate through the CNOT gates in \cref{fig:Color_Code_Review}. By propagating two-qubit Paulis at each CNOT location, we obtain the rules shown in \cref{tab:Ydecomp}. In particular, a rule of the form $P'_1P'_2 \to P_1P_2 \oplus P_3P_4$ means that, in \cref{Algo:DataGenOptimize}, the original fault $P'_1P'_2$ is replaced by two separately propagated components, $P_1P_2$ and $P_3P_4$. The resulting updates to \texttt{trainX} and \texttt{trainY} are then combined modulo two.

\subsection{Error propagation step to remove the presence of artificial error differences}
\label{subsecd:RemoveFakeErrorDiff}

\begin{figure}
     \centering
 \subfloat[\label{fig:Artificial_DiffEX} ]{\includegraphics[width=.5\textwidth]{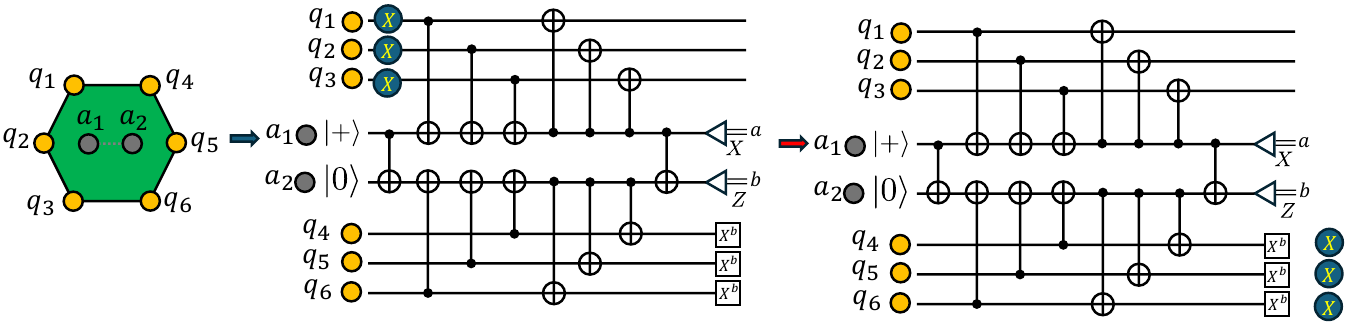}}
     \caption{ Example of an artificial error difference stored in \texttt{trainY} due to the feedforward operations used in the stabilizer measurement circuits. In this example, an input weight-three $X$ error with support on qubits $q_1$, $q_2$ and $q_3$ is mapped to the homologically equivalent weight-three error with support on qubits $q_4$, $q_5$ and $q_6$. }
     \label{fig:Artificial_Diff}
\end{figure}

Recall that the first two channels of \texttt{trainY} store error differences as described in \cref{eq:TrainY1,eq:TrainY2}. Due to the feedforward operations in the circuits shown in \cref{fig:Color_Code_Review}, artificial $X$ error differences can be obtained if the propagation of $X$ errors are not treated with care. An example is shown in \cref{fig:Artificial_Diff} where an input weight-three $X$ type error is mapped to a homologically equivalent output weight-three error with support on different qubits. In this case, an $X$ error difference would be recorded in \cref{eq:TrainY2} even though no new errors occurred, thus lowering the performance of our pre-decoders. Note that input errors being mapped to equivalent output errors with support on different qubits only occurs for $X$-type errors ($Z$-type errors are immune to this issue since the feedforward operation only affects $X$-type errors). 

\begin{algorithm}[H]
\caption{Error propagation protocol}
\begin{algorithmic}
\State Initialize $\tilde{e}^{0}_{\text{out}} = \mathbf{0}$.
\State Suppose there are $d_m$ rounds of syndrome measurements. 
 \For{$k = 1$ to $d_m - 1$}
     \State Let $E_k$ be the full set of faults generated by the noise model at each fault location in syndrome measurement round $k$, with potential appended errors that were undetected in the previous round as described in \cref{Algo:DataGenOptimize}.
     \State Let $\tilde{e}^{k}_{\text{temp}}$ be a binary vector for the output errors obtained by propagating $E_k$. 
     \State Compute $\tilde{e}^{k}_{\text{out}} = \tilde{e}^{k}_{\text{temp}} \oplus \tilde{e}^{k-1}_{\text{out}}$.
     \State Compute $X$ and $Z$ stabilizer syndromes using $\tilde{e}^{k}_{\text{out}}$.
     \State Error differences are computed in \cref{eq:XerrorDiff,eq:ZerrorDiff} by simply taking the $X$ and $Z$ components of $\tilde{e}^{k}_{\text{temp}}$.
 \EndFor
\end{algorithmic}
\label{Algo:DataStorage}
\end{algorithm}

In \cref{Algo:DataStorage}, we define an error-propagation protocol that explicitly tracks how faults evolve across rounds in order to prevent artificial differences in the resulting error configurations. This procedure is used in conjunction with \cref{Algo:DataGenOptimize}. The key idea is to avoid reintroducing errors from previous rounds as fresh input faults in the current round, which could otherwise be mapped to a different—though homologically equivalent—output error. Instead, feedforward corrections are computed using only the faults generated within the current round. Without this restriction, the weight-three input error in \cref{fig:Artificial_Diff} would incorrectly induce a feedforward correction that transforms the output into a weight-six error.

\subsection{Homological equivalence protocol for the color code}
\label{subsecd:ColorCodeHomological}

\begin{figure*}
     \centering
 \subfloat[\label{fig:HomoEquivAfterEX} ]{\includegraphics[width=.75\textwidth]{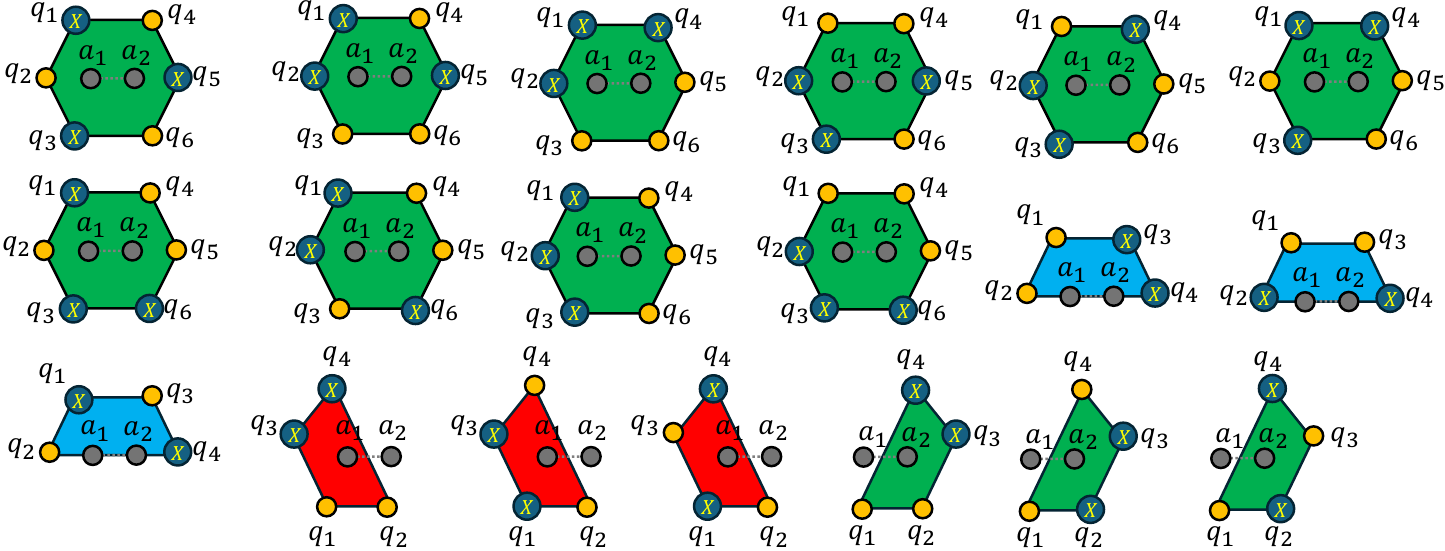}}
     \caption{ Configurations used in \texttt{fixEquivalenceX} for $X$ errors in the support of weight-6 and weight-4 $X$-type stabilizers. We pair each weight-3 $X$ error with its homologically equivalent error obtained by multiplying by the corresponding weight-6 stabilizer. The same rules are used for $Z$-type stabilizers.  }
     \label{fig:SpacelikeHomological}
\end{figure*}

When constructing the output labeled data in \texttt{trainY}, several errors (both spacelike and timelike) have equivalent representations. In particular, let $g \in \mathcal{S}$ where $\mathcal{S}$ is the stabilizer group of the color code. We say than an error $E'$ is homologically equivalent in the spacelike sense to an error $E$ if we can write $E' = g E$. As was shown in Refs.~\cite{Gicev2023scalablefast,ChambsLocalNN22,COLTI_Pre_surface}, choosing a fixed representation of homologically equivalent errors can significantly improve the training performance of our neural networks. Below we present a spacelike homological equivalence protocol for the color code which leads to greatly improved performance of our AI-based pre-decoders. We note that in Ref.~\cite{COLTI_Pre_surface} a timelike homological equivalence protocol was also developed. However we found numerically that combining a spacelike and timelike homological equivalence protocol for the color code led to inferior results compared to simply using the spacelike protocol presented below. 

\subsubsection{Spacelike homological equivalence protocol}
\label{subsubsec:spacelikeHomological}

When writing $E' = g E$, two things can happen. The hamming weight $|E'|$ can be reduced relative to $E$, or we can have $|E'| = |E|$ with errors shuffled along the support of the stabilizer $g$. If $g$ is an $X$ ($Z$) type stabilizer, applying $g$ to $E$ when $|E'| < |E|$ is referred to as \texttt{weightReductionX} (\texttt{weightReductionZ}), and when $|E'| = |E|$ as \texttt{fixEquivalenceX} (\texttt{fixEquivalenceZ}).

Let $g^{(X)}_6(j)$ correspond to the $j$-th weight-6 $X$-type stabilizer for the color code. Similarly, let $g^{(X)}_4(j)$ correspond to the $j$-th weight-4 $X$-type stabilizer for the color code. We define \texttt{weightReductionX} as follows. Let $d_6^{(X)}$ ($d_4^{(X)}$) be the number of weight-6 (weight-4) $X$-type stabilizers for a distance $d$ color code. For all $1 \le j_1 \le d_6^{(X)}$, we apply $g^{(X)}_6(j_1)$ to \texttt{trainY}. For all $1 \le j_2 \le d_4^{(X)}$, we apply $g^{(X)}_4(j_2)$ to \texttt{trainY}. If the number of 1's in \texttt{trainY} is reduced, we accept the change; otherwise, we leave \texttt{trainY} unchanged.

Next we describe the construction of \texttt{fixEquivalenceX}. For each weight-6 $X$-type stabilizer, we verify if a weight-3 $X$ error is in its support. If yes, we determine which configuration in \cref{fig:SpacelikeHomological} the error belongs to. If the configuration is not as in the ones shown, we modify the error in \texttt{trainY} to be the equivalent representation of one of the ten configurations shown. We perform an identical operation but for weight-2 $X$-type errors in the support of weight-4 $X$ stabilizers using the configurations shown in \cref{fig:SpacelikeHomological}.

Next define the function \texttt{simplifyX} which applies \texttt{weightReductionX} and \texttt{fixEquivalenceX} (in this order) to all $X$-type stabilizers of the color code. The function \texttt{simplifyX} is repeatedly applied until a steady state for \texttt{trainY} is achieved. By steady state, we mean that applying \texttt{simplifyX} to \texttt{trainY} no longer produces a change to \texttt{trainY}. 

We note that equivalent functions \texttt{weightReductionZ}, \texttt{fixEquivalenceZ} and \texttt{simplifyZ} can be used for $Z$ errors by simply replacing all $X$ operators with $Z$ operators given the properties of stabilizers for the color code. As such, the full spacelike homological equivalence function is executed by repeatedly applying \texttt{simplifyX} and \texttt{simplifyZ} until a steady state for \texttt{trainY} is achieved. 

Lastly, we remark that not all choices of \texttt{fixEquivalenceX} results in convergence when repeatedly applying \texttt{simplifyX}. That is, repeatedly applying \texttt{simplifyX} could result in errors continuously being shifted around without ever reaching a steady state. However, the choice of \texttt{fixEquivalenceX} in \cref{fig:SpacelikeHomological} does lead to convergence when repeatedly applying \texttt{simplifyX}.

\section{Numerical results}
\label{sec:Numerics}

\begin{table*}[t]
\centering
\footnotesize
\setlength{\tabcolsep}{4pt}
\renewcommand{\arraystretch}{0.95}
\begin{tabular}{|c|c|c|c|c|}
\hline
 & \texttt{num\_filters} & \texttt{kernel\_size} & RF size & \texttt{num\_params} \\
\hline
Model 1 & [128,128,128,4] & [3,3,3,3] & 9  & 912,272 \\
Model 2 & [256,256,256,4] & [3,3,3,3] & 9  & 3,595,012 \\
Model 3 & [128,128,128,4] & [5,5,5,5] & 17 & 4,224,388 \\
Model 4 & [128,128,128,128,128,4] & [3,3,3,3,3,3] & 13 & 1,797,764 \\
Model 5 & [256,256,256,256,256,4] & [3,3,3,3,3,3] & 13 & 7,134,468 \\
\hline
\end{tabular}
\caption{Pre-decoder models considered in this work. The size of the vectors used for \texttt{num\_filters} and \texttt{kernel\_size} indicate how many 3DConv layers are used. The entries in \texttt{num\_filters} and \texttt{kernel\_size} indicate the number of filters and kernel size used in that given layer. Note that if an entry in the $j$-th column of \texttt{kernel\_size} is $K$, a kernel size of $K \times K \times K$ is used in that layer. All models use stride 1 and no dilation.  }
\label{tab:models}
\end{table*}

In this section we present numerical results for the family of pre-decoder models trained using the fully convolutional neural networks summarized in \cref{tab:models} and illustrated in \cref{fig:ConvArchEX} as well as the model using five layers of Bottleneck blocks shown in \cref{fig:Arch_Cascade}. We refer to the model with Bottleneck layers as model B, which we open-source as a pre-trained checkpoint on \href{https://huggingface.co/collections/nvidia/nvidia-ising}{Hugging Face}. Note that model B has 2,936,580 parameters which is considerably less than model 5 despite having the same receptive field and $p_i=p_{out}=512$. Our model selection here prioritizes decoding speed: these are deliberately tiny models, and we expect significantly higher accuracy is achievable by scaling the decoder architectures, which we leave to separate work focused on accuracy. We further open-source our simulation code and training recipes on \href{https://github.com/nvidia/ising-decoding}{GitHub} to make these baselines reproducible and easy to customize.

\begin{table*}
\centering
\resizebox{\textwidth}{!}{%
\begin{tabular}{|c|c|}
\hline
\textbf{Hyperparameters} &  \textbf{Values} \\ \hline
Shots per epoch   &  16,777,216 \\ \hline
Number of epochs   & 400 \\ \hline
Batch size per GPU   & Epoch 1: 256, $\text{Epoch} \ge 2$: 1024  \\ \hline
Number of GPUs   & 4 \\ \hline
Optimizer   &  Lion: $\text{Weight decay} = 10^{-7}$, $\text{beta2} = 0.95$ \\ \hline
Learning rate schedule   & Warmup then decay (100 warmup steps). Apply $\gamma=0.7$ at milestones $[0.25,0.5,1.0]$  \\ \hline
Learning rates   & $\text{Model 1} = 2\times10^{-5}$, $\text{Model 2} = 1\times10^{-5}$, $\text{Model 4} = 1\times10^{-5}$, $\text{Model 5} = 1\times10^{-5}$  \\ \hline
Activation function   & GeLU (tanh approximation) \\ \hline
Dropout   & 0.01  \\ \hline
Exponential moving average (ema)   & $\text{decay} = 0.0001$  \\ \hline
Physical error rate & $\text{Model 1} = 0.003$, $\text{Others} = 0.004$ \\ \hline
 \end{tabular}
} 
\caption{ Hyperparameters used to train models 1 to 5 from \cref{tab:models}. The $\gamma=0.7$ is applied to the learning rate at milestones $[0.25,0.5,1.0]$. For instance, the first milestone 0.25 indicates that at $25\%$ of training steps, the learning rate becomes $0.7 \times \text{base}$. The tanh approximation of GeLU uses the function $\text{GeLU}(x) \approx 0.5 x (1 +\tanh{(\sqrt{2 / \pi}(x + 0.044715 x^3))})$.  }
\label{tab:hyperparameters}
\end{table*}

The pre-decoder models trained using the fully three-dimensional convolutional architecture use the hyperparameters listed in \cref{tab:hyperparameters} with data generated on the fly using a code distance $d$ matching each model's receptive field. Model B uses the same hyperparameters as in \cref{tab:hyperparameters} except that the learning rate is chosen to be $10^{-5}$ and the activation functions are SiLU instead of GeLU (see \cref{fig:Arch_Cascade}). Unless otherwise stated, simulations throughout this section employ the following depolarizing circuit-level noise model:

\begin{itemize}
\item A $|0\rangle$ ($|+\rangle$) state preparation is followed by an $X$ ($Z$) error with probability $2p/3$.
\item Prior to each $Z$ ($X$) basis measurement, an $X$ ($Z$) error occurs with probability $2p/3$.
\item With probability $p$, each two-qubit gate is followed by a two-qubit Pauli error drawn uniformly from $\{I,X,Y,Z\}^{\otimes 2} \setminus \{I\otimes I\}$.
\item During idle locations associated with either CNOT gates or state-preparation and measurement, a Pauli error is drawn uniformly from $\{X,Y,Z\}$ with probability $p$.
\end{itemize}

\begin{figure*}
    \centering
\subfloat[\label{fig:ler_vs_p_Model_1} ]{\includegraphics[width=.8\textwidth]{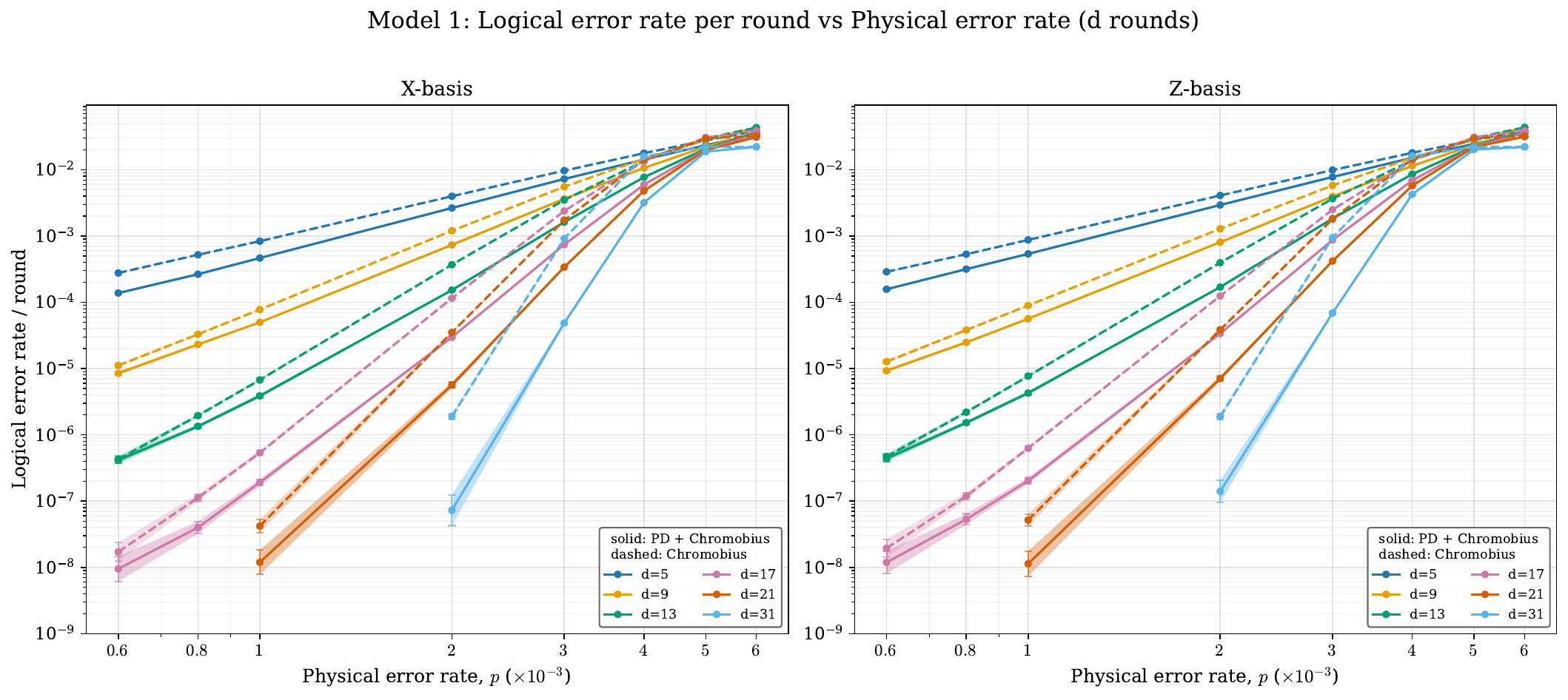}}
\vfill
\subfloat[\label{fig:ler_vs_p_Model_5} ]{\includegraphics[width=.8\textwidth]{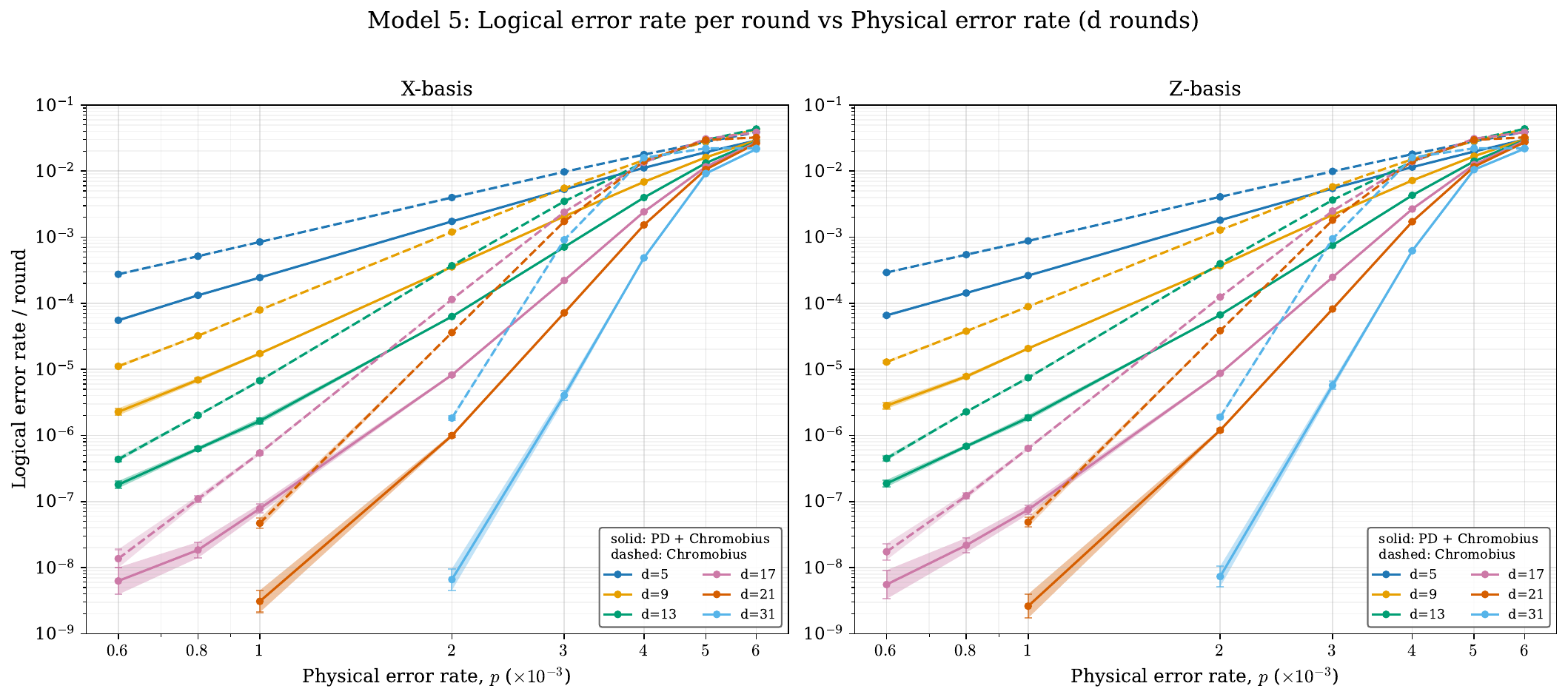}}
\caption{Plots of per-round LER for Chromobius (dashed lines) vs per-round LER of a pre-decoder model followed by Chromobius (solid lines). Per-round LERs were computed by calculating the LER for $d$ rounds and dividing the result by $d$.}
\label{fig:LERPlots}
\end{figure*}

\subsection{Pre-decoder logical error rates and syndrome densities with Chromobius as the global decoder}
\label{subsec:ChromobiusLER_SDR}

\begin{table}[htbp]
\centering
\begin{tabular}{|l|c|c|c|c|c|c|}
\hline
\textbf{Model} & $d=5$ & $d=9$ & $d=13$ & $d=17$ & $d=21$ & $d=31$ \\
\hline
Model 1 & $1.33 \text{x}$ & $1.51 \text{x}$ & $2.13 \text{x}$ & $3.18 \text{x}$ & $5.03 \text{x}$ & $18.58 \text{x}$ \\
Model 4 & $1.68 \text{x}$ & $2.23 \text{x}$ & $3.87 \text{x}$ & $7.67 \text{x}$ & $15.73 \text{x}$ & $124.71$x \\
Model 5 & $1.82 \text{x}$ & $2.64 \text{x}$ & $4.84 \text{x}$ & $10.61 \text{x}$ & $23.78 \text{x}$ & $223.72 x$ \\
Model B & $1.79 \text{x}$ & $2.88 \text{x}$ & $5.56 \text{x}$ & $13.16 \text{x}$ & $31.21 \text{x}$ & $\mathbf{347.74 \text{x}}$ \\
\hline
\end{tabular}
\caption{LER improvement factor ($X$-basis) for models 1, 4 and 5 of \cref{tab:models} followed by Chromobius compared to Chromobius alone. All data is obtained at $p=0.3 \%$.}
\label{tab:LER_Improvement}
\end{table}

\begin{figure*}
    \centering
\subfloat[\label{fig:sdr_vs_p_Model_1} ]{\includegraphics[width=.8\textwidth]{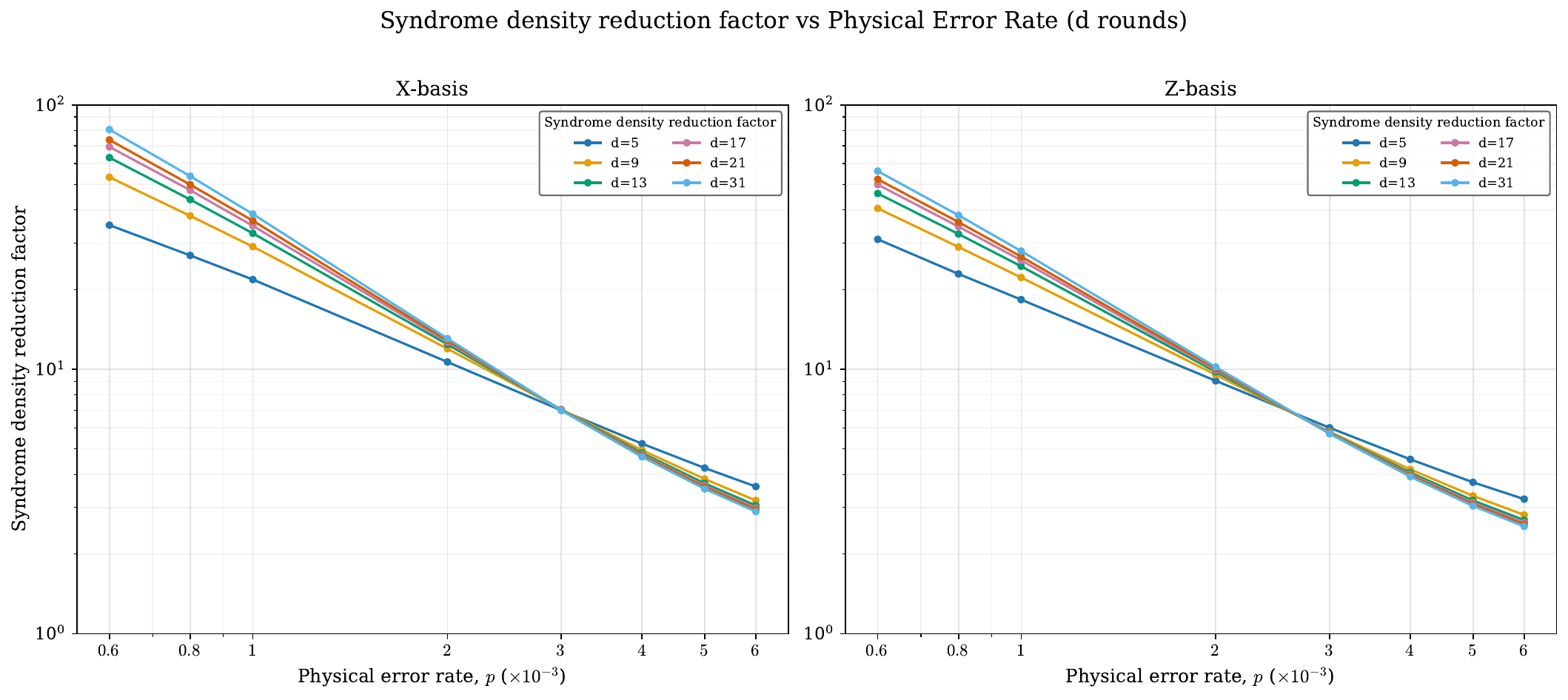}}
\vfill
\subfloat[\label{fig:sdr_vs_p_Model_5} ]{\includegraphics[width=.8\textwidth]{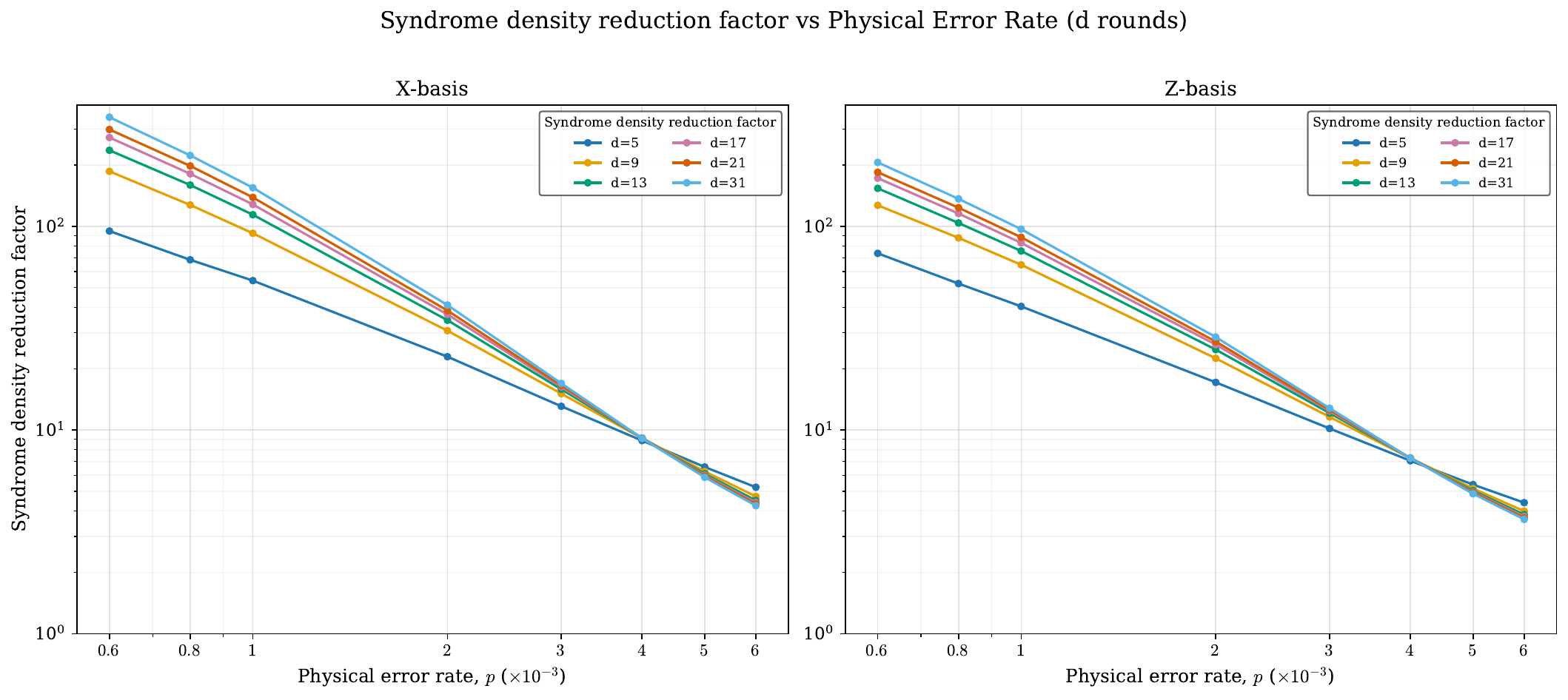}}
\caption{Plots of the syndrome density reduction factor for models 1 and 5 as a function of the physical error rate $p$ at various code distances. In (a) we show results for model 1 and in (b) for model 5.  }
\label{fig:SDRPlots}
\end{figure*}

Plots of the per-round logical error rates (LERs) for Chromobius and the pre-decoder + Chromobius pipeline are shown in \cref{fig:LERPlots} when using models 1 and 5 (results for model B are given in \cref{fig:Global_Runtime_Compare}). We illustrate results for model 1, which has the fastest inference time on a GPU, and model 5, which provides the best LER performance.

Several observations are immediately apparent. For both models, the pre-decoder + Chromobius pipeline achieves lower LERs than Chromobius alone across all sampled physical error rates, including $p=0.1 \%$, which is relevant for state-of-the-art hardware architectures. Most importantly, in nearly all cases, the relative LER improvement increases with code distance, even though the pre-decoder models were trained only at a fixed distance equal to the receptive-field size. These trends are quantified more explicitly in \cref{tab:LER_Improvement}. Note that in \cref{tab:LER_Improvement}, we also provide the LER improvement at $p=0.3 \%$ for model B which gives the best LER improvements across all considered models (although at the cost of slower runtimes as shown in \cref{subsec:Chromobius_Runtimes}).

The improvements are substantial. For example, at $p=0.3 \%$, the pre-decoder + Chromobius pipeline at $d=13$ achieves a lower LER than raw Chromobius decoding at $d=31$. This corresponds to obtaining superior logical performance using only 253 physical qubits (127 data qubits and 126 ancillas) instead of 1441 physical qubits (721 data qubits and 720 ancillas), while simultaneously achieving significantly lower decoding runtimes (see \cref{subsec:Chromobius_Runtimes}). We additionally observe an increase in the effective threshold when using pre-decoding prior to Chromobius.

This improvement can also be understood in terms of the effective distance scaling. For standalone Chromobius, which implements a Moebius decoder, the logical failure probability is expected to scale approximately as $p_L(d,p) \sim p^{\alpha d}$ with $\alpha \simeq 3/7$~\cite{PRXQuantum.3.010310}. Fitting the raw Chromobius total logical failure rates in our data gives $\alpha \approx 0.43$, in close agreement with this expectation. Repeating the same fit for the pre-decoder + Chromobius pipeline yields a larger effective exponent. For model 5, fitting through $p \le 0.3\%$ gives $\alpha \approx 0.49$ for both $X$- and $Z$-basis measurements, matching the scaling of \cite{Lee2025colorcodedecoder} and corresponding to an effective scaling closer to $p^{3.4d/7}$. This indicates that the pre-decoder improves not only the prefactor but also the observed distance scaling of the combined decoder. However, this scaling still does not saturate the optimal expectation $p_L(d,p) \sim p^{(d+1)/2}$. Thus, the pre-decoder moves the practical scaling closer to the optimal behavior, but does not fully reach it over the range of distances and physical error rates studied here.

\begin{figure*}
    \centering
\subfloat[\label{fig:physical_qubits_vs_ler_basis_X_p0p001} ]{\includegraphics[width=.45\textwidth]{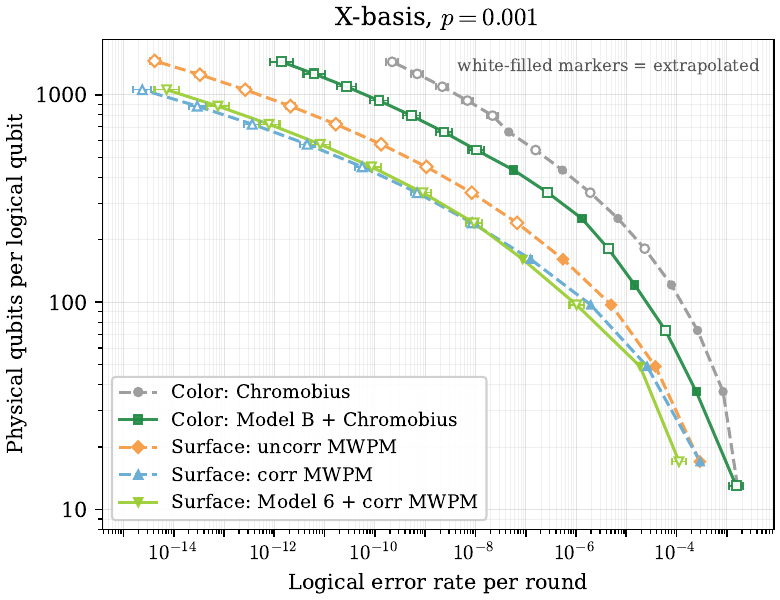}}
\subfloat[\label{fig:physical_qubits_vs_ler_basis_X_p0p003} ]{\includegraphics[width=.45\textwidth]{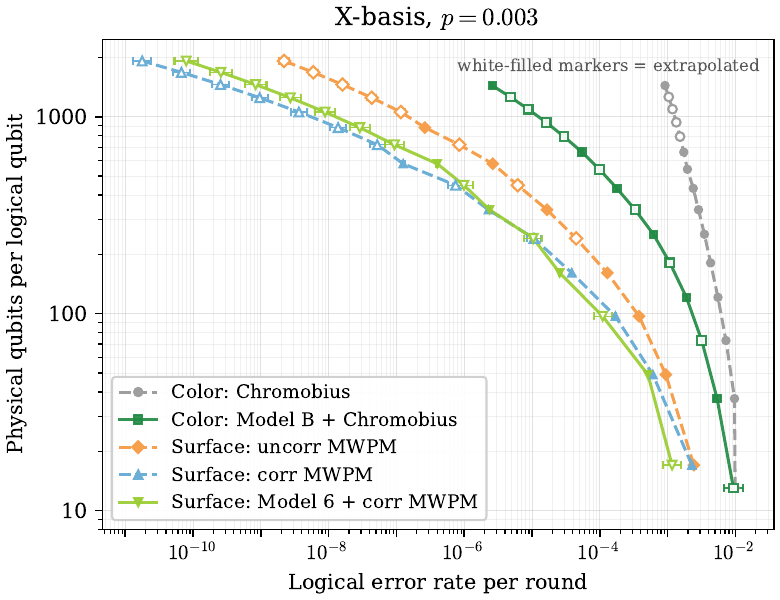}}
\caption{ Physical qubit count required to achieve a desired logical failure rate for both surface codes and color codes with various decoders. In (a), numbers are for physical error rates $p=0.001$ and in (b) for $p=0.003$.  }
\label{fig:Physical_Qubit_Count}
\end{figure*}

A distance-$d$ triangular color code using the syndrome-extraction circuits of \cref{fig:Color_Code_Review} has a total qubit count (data + ancilla) given by
\begin{align}
n_{\mathrm{color}} = \frac{3d^2-1}{2}.
\label{eq:n_color}
\end{align}
For comparison, the rotated surface code \cite{TomitaRotatedSurface2014} requires
\begin{align}
n_{\mathrm{surface}} = 2d^2-1.
\label{eq:n_surface}
\end{align}

\begin{table*}
\centering
 \resizebox{\textwidth}{!}{%
\begin{tabular}{|l|c|c|c|c|c|c|}
\hline
\textbf{Model} & \textbf{d=13, $p=0.001$ ($\mu$s/\text{round})} & \textbf{d=13, $p=0.003$ ($\mu$s/\text{round})} & \textbf{d=21, $p=0.001$ ($\mu$s/\text{round})} & \textbf{d=21, $p=0.003$ ($\mu$s/\text{round})} & \textbf{d=31, $p=0.001$ ($\mu$s/\text{round})} & \textbf{d=31, $p=0.003$ ($\mu$s/\text{round})} \\
\hline
Chromobius & 9.998 & 36.694 & 41.455 & 161.203 & 144.794 & 577.129 \\
Chromobius after model 1 (GeLU) & 1.739 & 11.527 & 6.965 & 44.600 & 23.360 & 156.279 \\
Chromobius after model 4 (GeLU) & 1.242 & 7.484 & 5.466 & 26.314 & 18.530 & 87.312 \\
Chromobius after model 5 (GeLU) & 1.050 & 6.321 & 4.450 & 22.424 & 15.718 & 72.987 \\
Chromobius after model B & 1.058 & 6.125 & 4.607 & 20.836 & 16.125 & 64.268 \\
Pre-decoder model 1 (GeLU) & 3.44 & 3.44 & 3.19 & 3.19 & 3.73 & 3.73 \\
Pre-decoder model 4 (GeLU) & 4.23 & 4.23 & 4.08 & 4.08 & 4.94 & 4.94 \\
Pre-decoder model 5 (GeLU) & 6.12 & 6.12 & 6.52 & 6.52 & 9.17 & 9.17 \\
Pre-decoder model B (SiLU) & 8.74 & 8.74 & 9.44 & 9.44 & 14.39 & 14.39 \\
\hline
\end{tabular}
}
\caption{Comparison of runtimes for Chromobius (both with and without syndromes processed by pre-decoder models) and pre-decoder models. All results correspond to the task of decoding a single (batch size $=1$) $d\times (d + (d-1)/2)\times d$ block, and we report averaged runtimes per syndrome measurement round. Chromobius runtimes are computed using a Grace Neoverse-V2 CPU. The label ``Chromobius after model $X$'' refers to Chromobius runtimes after processing syndromes by the pre-decoder model $X$ (i.e. one of the 5 models in \cref{tab:models}). GPU runtimes for all five pre-decoder models are computed using an NVIDIA GB300 GPU using TensorRT with FP8 precision. All results are for $X$-basis measurements. }
\label{tab:runtimes_chromobius}
\end{table*}

In \cref{fig:Physical_Qubit_Count}, we plot the number of physical qubits, computed using \cref{eq:n_color,eq:n_surface}, required to reach a specified target logical error rate for both surface codes and color codes. We show results for several decoders. The figure demonstrates that, for color-code memories, the model B + Chromobius decoder substantially reduces the required number of physical qubits relative to Chromobius alone. However, for pure quantum-memory benchmarks, \cref{fig:Physical_Qubit_Count} also shows that surface codes can reach a given target logical error rate using fewer physical qubits than color codes, even without the use of a pre-decoder.

Nevertheless, logical operations implemented using transversal gates and lattice surgery can be substantially more efficient for color codes than for surface codes. Thus, these results suggest that AI-based pre-decoding may significantly improve the competitiveness of color codes for universal fault-tolerant quantum computation, even if surface codes retain an advantage in the pure-memory setting. The qubit overhead of color codes could potentially be reduced further by using the “middle-out” syndrome-extraction circuits of Ref.~\cite{GidneyColor2023}, which avoid dedicated ancilla qubits. Additional exploration of model architectures and training strategies may further narrow the remaining qubit-count gap between surface codes and color codes.

Finally, syndrome-density reduction (SDR) plots for models 1 and 5 are shown in \cref{fig:SDRPlots}. For $p \lesssim 0.1 \%$, SDRs exceeding two orders of magnitude are achieved, meaning that more than 99\% of syndromes have been canceled and resulting in substantial Chromobius runtime reductions (see \cref{subsec:Chromobius_Runtimes} for exact runtimes). Similar to the LER results, SDR performance generally improves with increasing code distance.

\subsection{Pre-decoder logical runtimes with Chromobius as the global decoder}
\label{subsec:Chromobius_Runtimes}

\begin{table}
\centering
\begin{tabular}{|l|c|c|}
\hline
$d$ & X-basis & Z-basis \\
\hline
5  & 0.351  & 0.336 \\
9  & 0.464  & 0.352 \\
13 & 1.050  & 0.621 \\
17 & 2.227  & 1.203 \\
21 & 4.450  & 2.180 \\
31 & 15.718 & 8.187 \\
\hline
\end{tabular}
\caption{Chromobius runtimes after applying Model 5 at $p=0.1$\%. Runtimes are reported in $\mu$s per round and are separated by measurement basis.}
\label{tab:runtimes_chromobius_asymmetry}
\end{table}

\begin{figure*}
    \centering
\subfloat[\label{fig:chromobius_runtime_bars_logEX} ]{\includegraphics[width=.9\textwidth]{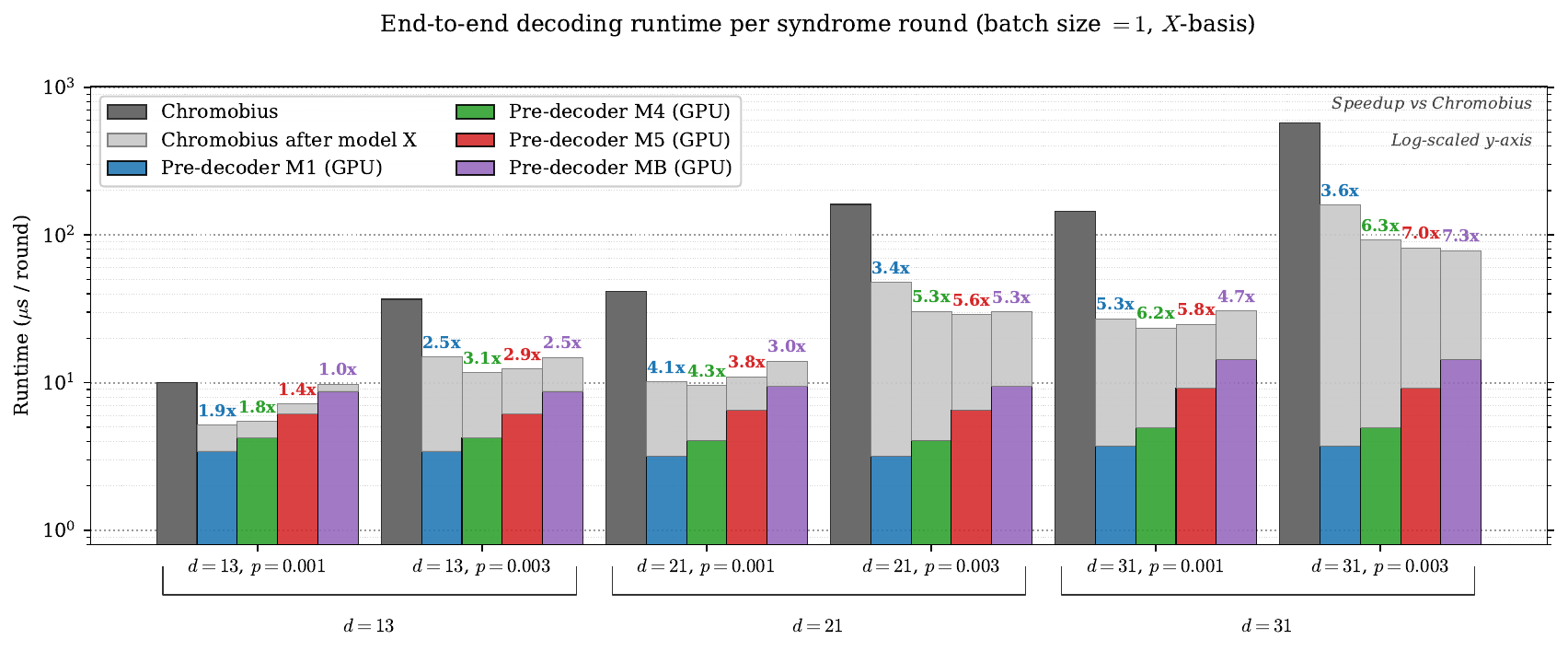}}
\caption{Total speedup factors when using a pre-decoder (model MX with GeLU activation) + Chromobius compared to Chromobius alone. The speedup is defined as the ratio between raw Chromobius runtimes and the sum of pre-decoder inference runtimes plus Chromobius runtimes after pre-decoding, averaged over both bases (see \cref{tab:runtimes_chromobius} and \cref{tab:runtimes_chromobius_asymmetry}). The largest speedup factor for each input setting is shown in bold.}
\label{fig:chromobius_runtime_bars_log}
\end{figure*}

\begin{figure*}
    \centering
\subfloat[\label{fig:end_to_end_runtime_vs_ler_basis_X_p0p001} ]{\includegraphics[width=.45\textwidth]{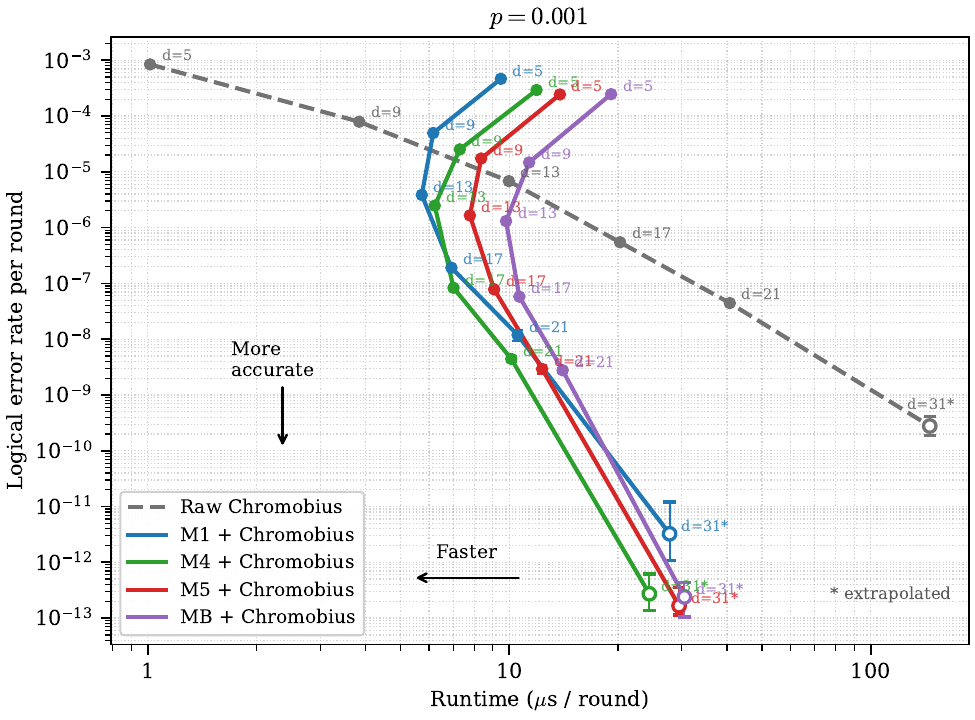}}
\subfloat[\label{fig:end_to_end_runtime_vs_ler_basis_X_p0p003} ]{\includegraphics[width=.45\textwidth]{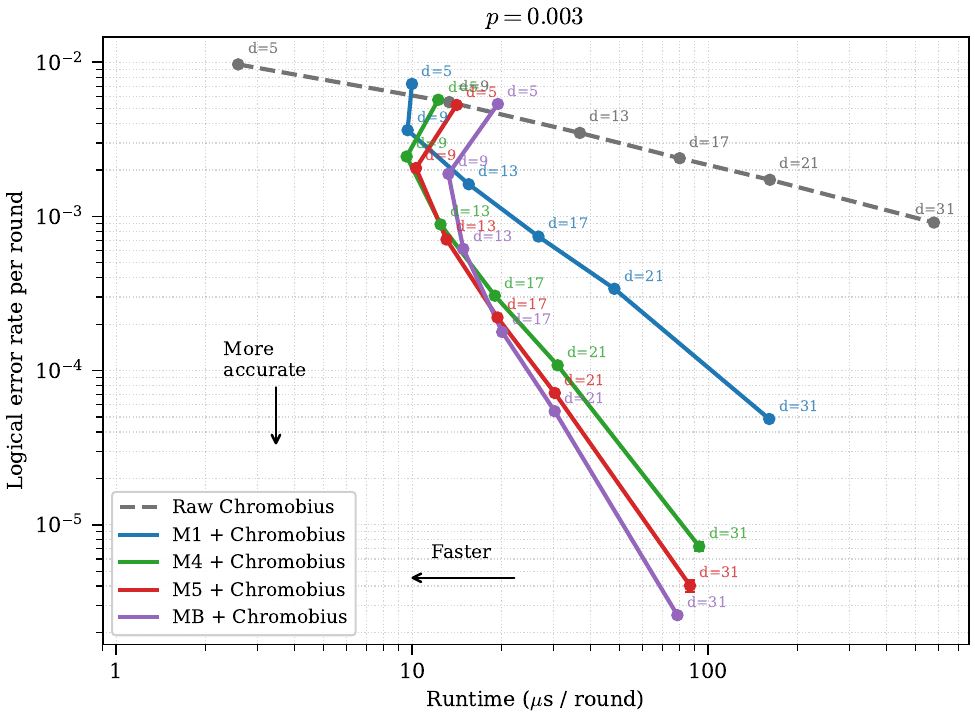}}
\caption{End-to-end per-round LERs and single-shot (batch size 1) runtimes for Chromobius and pre-decoder models + Chromobius with representative physical error rates $p=0.001$ (left) and $p=0.003$ (right) for X-basis. Pre-decoder models M1, M4, M5 and MB run at FP8 precision and were timed on a single GB300 GPU, while Chromobius was timed on a single Grace Neoverse-V2 CPU. }
\label{fig:Global_Runtime_Compare}
\end{figure*}

In this section, we characterize the runtimes of Chromobius both with and without pre-decoding, as well as the runtimes of the pre-decoder models themselves across various code distances and physical error rates. We then quantify the overall runtime reduction achieved by the pre-decoder + Chromobius pipeline relative to standalone Chromobius decoding.

\cref{tab:runtimes_chromobius} summarizes the runtimes of Chromobius, the pre-decoder + Chromobius pipeline, and the pre-decoder models 1, 4, 5 and B for $d=13,21$ and $31$ at physical error rates $p=0.1\%$ and $p=0.3\%$ for X-basis memory experiments. At $p=0.1 \%$, Model 5 yields the largest reductions in Chromobius runtimes, achieving up to a 9.5x improvement at $d=13$. At $p=0.3\%$, model B yields the largest reductions in Chromobius runtimes (a 9x improvement at $d=31$). However, model B is also the slowest of the considered pre-decoders (even though it uses 2,936,580 parameters compared to 7,134,468 for model 5), while model 1 provides the lowest pre-decoder latency. We focus on $p=0.1\%$ and $p=0.3\%$ since these are relevant error rates for near-term hardware architectures.

We observe a consistent Chromobius runtime asymmetry that is basis-dependent and that gets larger with decreasing physical error rates and increasing code distances. To exemplify it, \cref{tab:runtimes_chromobius_asymmetry} shows averaged Chromobius runtimes after applying Model 5 at $p=0.1$\% for both $X$ and $Z$ bases. While the asymmetry is barely noticeable at $d=5$, it grows to almost a 2x difference between X and Z-basis decoding runtimes at $d=31$. As observed in Ref.~\cite{GidneyColor2023}, the Chromobius decoder when used with the superdense syndrome extraction circuits given in \cref{fig:Color_Code_Review} results in asymmetric LERs between the $X$ and $Z$ basis. As such it should not be surprising to see these results amplified when using our pre-decoders given the improvement in LERs and runtimes.

In \cref{fig:chromobius_runtime_bars_log}, we summarize the total end-to-end runtime improvements achieved by the pre-decoder + Chromobius pipeline for models 1, 4, 5 and B. The largest observed overall speedup is 7.33x, obtained using model B at $d=31$ and $p=0.3\%$. All four models exhibit parameter regimes in which they outperform the others. Importantly, the overall speedup generally increases with code distance. For fixed code distance, larger speedups are typically observed at higher physical error rates due to the increased syndrome density, which significantly slows Chromobius decoding. In contrast, pre-decoder runtimes are completely insensitive to the physical error rate.

In \cref{fig:Global_Runtime_Compare}, we plot the end-to-end per-round LERs as a function of total decoding runtime for Chromobius and the pre-decoder + Chromobius pipelines for $X$-basis measurements. These plots illustrate the tradeoff between logical performance and decoding latency. For example, in \cref{fig:end_to_end_runtime_vs_ler_basis_X_p0p001} which shows data at $p=0.1 \%$, pre-decoding improves LERs for all considered distances, but for $d \le 9$, the total runtime remains larger than that of standalone Chromobius decoding. However, for $d \ge 13$, all four considered pre-decoder models simultaneously improve both LER and runtime relative to Chromobius alone. At $p=0.3 \%$, the simultaneous runtime and LER improvement regime extends to $d \ge 9$ (except for model B which offers an advantage for $d \ge 13$), as shown in \cref{fig:end_to_end_runtime_vs_ler_basis_X_p0p003}.

The $d=31$, $p=0.1\%$ logical error rates shown in \cref{fig:end_to_end_runtime_vs_ler_basis_X_p0p001} are extrapolated. For standalone Chromobius, we use the scaling reported for the Moebius decoder, $p_L(d,p) \propto p^{3d/7}$~\cite{PRXQuantum.3.010310}, applied to the measured total logical failure rates at $d=31$ and $p=0.2\%,0.3\%$. For the pre-decoder curves, we use an analogous local-slope extrapolation. We assume that for $d$ larger than the receptive field of the corresponding model the effective exponent per unit distance is approximately stable. We therefore estimate local power-law slopes from measured total LERs at post-receptive-field distances, rescale the inferred exponent density to $d=31$, and extrapolate the measured $d=31$ data down to $p=0.1\%$. These extrapolated points should be interpreted as order-of-magnitude estimates rather than direct measurements.
 
\section{Conclusion}
\label{sec:Conclusion}

In this work, we introduced an AI-based pre-decoder for triangular color codes. We developed a novel architecture for mapping color-code stabilizers onto a two-dimensional grid suitable for convolutional neural-network processing. We additionally introduced several data-processing techniques that substantially improve training performance, particularly for the complex syndrome-extraction circuits containing feedforward operations.

Using Chromobius as the global decoder, we demonstrated that the full pre-decoder + Chromobius pipeline achieves substantial reductions in both logical error rates (LERs) and decoding runtimes relative to standalone Chromobius decoding. Importantly, these gains generally become more pronounced as the code distance increases across nearly all sampled physical error rates. For example, at $d=31$ and $p = 0.3\%$, our pipeline achieves a 347x reduction in LER together with a 7.33x runtime improvement relative to Chromobius alone.

Several important directions for future work remain. First, broader model exploration may yield further improvements in both decoding performance and runtime. For example, incorporating projection layers between convolutional layers could reduce parameter counts while simultaneously improving model expressivity and performance \cite{7780459,Cascade_decoder}. Additional GPU-level optimizations, including the use of lower-precision arithmetic such as FP4, may further reduce pre-decoder inference latency, particularly on next-generation hardware architectures designed for low-precision computation. Further, current work is being done to develop a parallel implementation of pre-decoders on GPUs. Early results show substantial improvements in both throughput and latency compared to decoding entire volumes resulting in runtimes well below $1 \mu \text{s}$.

Finally, extending AI-based pre-decoders to fully support lattice-surgery operations within parallel block-wise decoding frameworks in both space and time will be crucial for scalable real-time fault-tolerant quantum computation. We believe that the results presented in this work provide further evidence that AI-assisted decoding may significantly improve the practicality and competitiveness of color-code architectures for universal fault-tolerant quantum computing.

\newpage 
\appendix

\bibliography{q}
 
\end{document}